\documentclass[12pt]{article}
\newcommand{\beq}{\begin{equation}}
\newcommand{\eeq}{\end{equation}}
\newcommand{\beqa}{\begin{eqnarray}}
\newcommand{\eeqa}{\end{eqnarray}}
\newcommand{\f}{\frac}

\newcommand{\sn}{{\rm sn}}
\newcommand{\cn}{{\rm cn}}
\newcommand{\dn}{{\rm dn}}
\newcommand{\ds}{{\rm ds}}
\newcommand{\cs}{{\rm cs}}
\newcommand{\ns}{{\rm ns}}
\newcommand{\cd}{{\rm cd}}
\newcommand{\nd}{{\rm nd}}
\newcommand{\sd}{{\rm sd}}
\newcommand{\nc}{{\rm nc}}
\newcommand{\dc}{{\rm dc}}
\newcommand{\ti}{\tilde}
\newcommand{\sech}{{\rm sech}}
\newcommand{\sss}{{\vspace{.2in}}}

\begin{document}
\begin{center}
{\LARGE {\bf  Cyclic Identities Involving Jacobi Elliptic
Functions. II}}
\end{center}
\vspace{.5in}
\begin{center}
{\Large{\bf  \mbox{Avinash Khare} }}\\
 \noindent {\large
Institute of Physics, Sachivalaya Marg, Bhubaneswar 751005, India}
\\
{\bf(khare@iopb.res.in)}
\end{center}
\vspace{.2in}
\begin{center}
{\Large{\bf  \mbox{Arul Lakshminarayan} }}\\ \noindent {\large
Physical Research Laboratory, Navrangpura, Ahmedabad 380009,
India}\\ {\bf (arul@prl.ernet.in)}
\end{center}
\vspace{.2in}
\begin{center}
{\Large{\bf  \mbox{Uday Sukhatme} }}\\ \noindent {\large
Department of Physics, State University of New York at Buffalo,
Buffalo, NY 14260-4600}\\ {\bf (sukhatme@buffalo.edu)}
\end{center}
\vspace{1.2in}
{\bf {Abstract:}} Identities involving cyclic sums of terms
composed from Jacobi elliptic functions evaluated at $p$ equally
shifted points on the real axis were recently found. These
identities played a crucial role in discovering linear
superposition solutions of a large number of important nonlinear
equations. We derive four master identities, from which the
identities discussed earlier are derivable as special cases.
Master identities are also obtained which lead to cyclic
identities with alternating signs. We discuss an extension of our
results to pure imaginary and complex shifts as well as to the
ratio of Jacobi theta functions.

\newpage
\section{Introduction}

In a recent paper \cite{ksjmp}, we have
given many new mathematical identities involving the Jacobi elliptic functions
$\sn \,(x,m)$, $\cn \,(x,m)$, $\dn \,(x,m)$, where $m$ is the elliptic
modulus parameter $( 0\leq m\leq 1)$. The
functions $\sn \,(x,m)$, $\cn \,(x,m)$, $\dn \,(x,m)$ are doubly periodic
functions with periods $(4K(m), i2K'(m))$, $(4K(m), i4K'(m))$,
$(2K(m), i4K'(m))$,
respectively \cite{abr}.
Here, $K(m)$ denotes the complete elliptic integral of the first kind, and
$K'(m)= K(1-m)$. The $m =0$ limit gives
$K(0)=\pi /2$ and trigonometric functions: $\sn(x,0)=\sin x, ~\cn(x,0)=\cos x,
~\dn(x,0)=1$. The $m \rightarrow 1$ limit gives $K(1) \rightarrow \infty$ and
hyperbolic functions: $\sn(x,1) \rightarrow \tanh x,
~\cn(x,1) \rightarrow \sech \,x, ~\dn(x,1) \rightarrow \sech\, x$. For simplicity, from now on
we will not explicitly display the modulus parameter $m$ as an argument of the
Jacobi elliptic functions.

The cyclic identities
discussed in ref. \cite{ksjmp} play an important role in showing
that a kind of linear superposition is valid for many nonlinear differential
equations of physical interest \cite{ksprl,cks}. In all identities, the
arguments of the Jacobi functions in successive terms
are separated by either $2K(m)/p$ or $4K(m)/p$, where $p$ is an integer.
Each $p$-point
identity of rank $r$ involves a cyclic homogeneous polynomial of degree
$r$ (in Jacobi elliptic functions with $p$ equally spaced arguments) related
to other cyclic homogeneous  polynomials of degree $r-2$ or smaller.
In ref. \cite{ksjmp}, explicit algebraic proofs were given for specific small values of $p$
and $r$ by
using standard properties of Jacobi elliptic functions. However, identities
corresponding to higher values of $p$ and $r$ were only verified numerically
using advanced mathematical software packages. In this article, we present rigorous
mathematical proofs valid for arbitrary $p$ and $r$. As a useful byproduct, we
determine explicit forms for the constants appearing in various identities.
All the identities in ref. \cite{ksjmp} corresponded to real shifts of multiples of
$2K(m)/p$ or $4K(m)/p$. Here, we discuss how to
obtain new identities corresponding to pure imaginary shifts by
multiples of $i2K'(m)/p$ or $i4K'(m)/p$, as well as identities
corresponding to complex shifts by multiples of $2[K(m)+iK'(m)]/p$
or $4[K(m)+iK'(m)]/p$. We also discuss the identities for the nine
secondary Jacobi elliptic functions like
$~\cd \,(x,m)$, $~\ns \,(x,m)$, $~\ds \,(x,m)$.
Also, we give results for several identities involving Weierstrass elliptic functions and ratios of Jacobi
theta functions, both of which are
intimately related with the Jacobi elliptic functions \cite{abr}.
In our proofs, we classify the identities
into four types, each with its own ``master identity'' which we
prove using a combination of the Poisson summation formula and the
special properties of elliptic functions.

All our identities involve sums of the following generic form:
\beq \label{sum}
S_p(x_0)=\sum_{j=1}^{p} f(x_j)~,
\eeq
where $f(x)$ is composed from Jacobi elliptic functions
with arguments
corresponding to $p$ equally spaced points \[
x_j=x_0+(j-1)T/p, \;\;j=1,\dots,p~~,
\]
where $T$ is a
period of $f(x)$ and the base point $x_0$ is an arbitrary
complex number.
We define the quantities $P,Q$ by:
\beq f(z + 2i K')= (-1)^P f(z),\;\; f(z+2K)=(-1)^Q f(z), \;\;
P,Q=0,1~.
\eeq
Note that $Q=0,1$ correspond to real periods $2K(m),4K(m)$ and $P=0,1$
correspond to pure imaginary periods $i2K'(m),i4K'(m)$ respectively.
We denote the four possibilities as $(+,+)$, $(-,+)$,
$(+,-)$ and $(-,-)$, where the first sign refers to the sign of
$(-1)^P$ and the second to that of $(-1)^Q$. We will
derive master identities for each of these four possibilities.

For example, one of the simplest identities discussed in ref.
\cite{ksjmp} reads \beq \label{ddA}
\sum_{j=1}^{p}\dn(x_j)\dn(x_{j+1})=A~, \eeq where $A$ is a
constant independent of the base point $x_0$, $T=2K$ and $p$ is
any integer. In this case, we have $f(z)=\dn(z)\, \dn(z+T/p)$
which corresponds to $P=0,Q=0$, since $\dn(z+2K)=\dn(z)$ and
$\dn(z+2iK')=-\dn(z)$. Liouville's theorem can be used to prove
the above identity, since $\dn(z)$ has simple poles within its
fundamental region $(0,2K,2K+4iK',4iK')$  at $iK'$ and $3iK'$ both
of which we collectively refer to as $z^*$ . The identity
$\dn(z^*+u)+\dn(z^*-u)=0$ for arbitrary complex $u$ then implies
that every pole in the sum is cancelled exactly by a zero of the
same order. Thus the sum is an analytic function without any poles
in the finite part of the complex plane and by Liouville's theorem
must be a constant \cite{nev}. This is an explicit illustration of
the general principle underlying the identities, namely that the
orders of poles in a higher order polynomial are reduced by some
zeros leading to simpler sums. However this method does not yield
the constants like $A$ explicitly. In fact, using the Poisson
summation formula and special properties of Jacobi elliptic
functions, we show below that the constant $A$ in Eq. (\ref{ddA})
is given by
\beq \label{ddB} A=\f{p}{2K}\int_{0}^{2K} \dn(x) \dn(x+T/p)\, dx
=p\,\left( \dn(2K/p) -\f{\cn(2K/p)
Z(\beta_{2K})}{\sn(2K/p)}\right)~, \eeq where $Z$ is the Jacobi
zeta function ($Z \equiv Z(\beta_q,m)$) \cite{abr} with
$\beta_{q}\equiv\arcsin(\sn(q/p,m))$ being the Jacobi Amplitude
function.

{}Identities analogous to Eq. (\ref{ddA}) also hold for $\sn$ and $\cn$. For instance,
\beq \label{ss}
\sum_{j=1}^{p}\sn(x_j)\,\sn(x_{j+1})=\f{p~Z(\beta_{2K})}{m~\sn(2K/p)}~.
\eeq Here $T$ is $2K$ as this is the periodicity of $f(z)=\sn(z)
\,\sn(z+T/p)$. The expression as given above is valid for all
integer values of $p>2$, with both sides vanishing when $p=2$.

A further generalization that can be easily treated with the
techniques developed below is to sums that involve $r$-th
neighbors. We consider the case when $r$ and $p$ are coprime
integers and $1 \le r <{p-1}$, the other cases being included
since identities for any choice of $p$ also include the identities
for the factors of $p$. Such a generalization of say Eq.
(\ref{ddB}) above is \beq \label{ddr}
\sum_{j=1}^{p}\dn(x_j)\dn(x_{j+r}) =p\,\left( \dn(2rK/p)
-\f{\cn(2rK/p) Z(\beta_{2rK})}{\sn(2rK/p)}\right)~. \eeq Another
easy generalization is to identities involving terms consisting of
a product of an arbitrary number of Jacobi elliptic functions.

The plan of this article is as follows. In Sec. 2, we derive the master
identities which form the basis for obtaining all the identities in this
paper, and are tabulated in Appendices A and B.  Sec. 3 contains a
derivation of identities involving alternating signs. In Sec. 4, we present a
collection of comments, including some which permit a generalization of all
the identities to incorporate pure imaginary and complex shifts
and to present the identities for Weierstrass functions as well as for
ratios of Jacobi theta functions.

\section{The Master Identities}

In this section, we derive the four master identities
corresponding to $Q,P$ taking on values $0,1$, which effectively encompass
most of the cyclic identities discussed in ref. \cite{ksjmp}. The
remaining identities in ref. \cite{ksjmp} correspond to master
identities with alternating signs, and these are discussed in the next section.

For completeness we first derive the finite version of the Poisson
summation formula \cite{Poissonform} that fully exploits the
equally spaced nature of the sampling points, and which plays a
crucial role in subsequent derivations. Since $f(x)$ has a period $T$, we may
expand it in a Fourier series:
\beq \label{poi3}
f(x)=\f{1}{T}\sum_{k=-\infty}^{\infty} a_k\, e(k x/T) ~,~~
a_k= \int_0^{T} f(x) \,e\left(- k
x/T\right) \, dx~,
\eeq
where we have introduced the convenient notation $e(x) \equiv \exp(2 \pi i x)$.
The required sum may then be written as:
 \beq
\label{poi1} S_p(x_0)=\sum_{j=1}^{p}
f(x_j)=\f{1}{T}\sum_{k=-\infty}^{\infty} a_k\,
e(kx_0/T)\,\sum_{j=1}^{p} e(kj/p). \eeq
 Using the simple identity
\beq \sum_{j=1}^{p}\, e(kj/p)=\left\{ \begin{array}{ll} p
&\mbox{if}\;\; p|k\\0 &\mbox{otherwise~,}\end{array}\right.
\eeq
we
get
\beq \label{poisson} S_p(x_0)= \frac{p}{T} \,\sum_{k,\, p|k}
a_k \,e\left(k x_0/T\right)~.
 \eeq
Note that we need to evaluate only those Fourier coefficients $a_k$ for which $k$ is a
multiple of $p$.

\subsection{ Cases corresponding to $Q=0$}

We first derive the two master identities corresponding to
$Q=0$ [or equivalently $T=2K$], allowing $P$ to be either $1$
or $0$. Consider the rectangle $ABCD \equiv
(-K,K,K+2iK',-K+2iK')$. We assume that $f(z)$ has  a
finite number of poles inside $ABCD$ situated at points $z^{\ast}_w=iK'
+ wT/p$, where $w=0,\pm 1,\pm 2,...$ and $|w|<p$. Let the
principal part of $f(z)$ about the pole $z^{\ast}_w$ be
\begin{equation}
\sum_{l=1}^{L_w} \f{\alpha_l^{(w)}}{(z-z^{\ast}_w)^l}~,
\end{equation}
which makes this a pole of order $L_w$.

We now use the fact that $f(z)$ is composed of elliptic functions, and this
essentially allows evaluation of $a_k$. To evaluate $a_k$, for
$k\ne0$, consider the integral over the rectangle $ABCD$: \beqa &&\oint
f(z)\,e(-k z/T)\, dz = a_k + \int_{K+2iK'}^{-K+2iK'}f(z)\, e(- k
z/T)\, dz \nonumber
\\ &&= a_k+\int_{K}^{-K}f(z+2iK')\, e(-k [z+2iK']/T)\,
dz \nonumber
\\ &=& a_k+(-1)^{P+1} q^{-2k} \int_{-K}^{K}f(z)\, e(-k z/T)\,dz
= a_k[1+(-1)^{P+1} q^{-2k}]~, \eeqa where $q= \exp(-\pi K'/K)$ is
the Jacobian nome \cite{abr}. The contributions of the vertical
segments of the integration contour are equal and opposite, and
cancel each other.

On the other hand, the sum of the residues of $f(z) \,e(-k z/T)$
may also be calculated. The residue at the pole $z^{\ast}_w$ is:
\beqa
&&\mbox{Res}\left[ f(z) \,e(-k z/T)\right]
=\mbox{Res}\left[ f(z) \,e\left(-k [z-(iK' + wT/p)]/T\right)
q^{-k}\,e(-k w/p)\right] \nonumber
\\ &&=q^{-k}\mbox{Res}
\left[\left(\sum_{l=1}^{L_w}\frac{\alpha_l^{(w)}}{(z-z^{\ast}_w)^l}
\right) \left(\sum_{n=0}^{\infty} (\frac{-2 \pi i k}{T})^n
\frac{(z-z^{\ast}_w)^n}{n!}\right) \right] \nonumber \\
&&=q^{-k}\sum_{l=1}^{L_w} \left[
\frac{\alpha_l^{(w)}}{(l-1)!}\left(\frac{-2 \pi i
}{T}\right)^{l-1} k^{l-1} \right]~.
\eeqa
For the second equality,
we have made use of the fact that only those
$a_k$ for which $k/p$ is an integer need to be evaluated, thanks to the Poisson
summation formula Eq.~(\ref{poisson}).

Define $L'\equiv \mbox{Max}\{L_1, L_2, \ldots, L_w\}$ and
$\gamma_l= \sum_{w} \alpha_l^{(w)} \; ,~l=1,\ldots,L'$, where we set
nonexistent $\alpha^{(w)}$ to be zero. We also set $L$ to be the
maximum integer such that $\gamma_L$ is nonzero. If $L=0$, there
are no nonvanishing $\gamma$ and the function is regular. Using
$T=2K$ the sum of the residues at all the interior poles may be
written as: \beq \mbox{Res}=q^{-k}\sum_{l=1}^{L} \left[
\f{\gamma_l}{(l-1)!}\left(\f{-2 \pi i }{2K}\right)^{l-1} k^{l-1}
\right]~. \eeq Thus
\begin{equation}\label{ordak1}
a_k=\f{2 \pi i q^{-k}}{1+(-1)^{P+1}q^{-2k}} \sum_{l=1}^{L} \left[
\f{\gamma_l}{(l-1)!}\left(\f{-2 \pi i }{2K}\right)^{l-1} k^{l-1}
\right]~,~~k \ne 0~.
\end{equation}
Therefore, Eq. (\ref{poisson}) now becomes \beq \label{mother1}
S_p(x_0)= \f{p}{2K} \left[  a_0 + 2\pi i  \sum_{l=1}^{L}
\frac{\gamma_l}{(l-1)!}\left(\frac{-2 \pi i }{2K}\right)^{l-1}
\sum_{k\ne0, \,p|k} \frac{k^{l-1} q^{-k} }{1+(-1)^{P+1}q^{-2k}}\,
 e\left(\f{k x_0}{2K}\right) \right]~.
\eeq

We are in a position to derive two master identities (MI)
corresponding to the $P=1$ and $P=0$ cases, which we call MI~-~I
and MI~-~II respectively. We state for convenience the following
well-known symmetry properties:
\[ \sn(z+2K)=-\sn(z), \, ~\cn(z+2K)=-\cn(z), ~\, \dn(z+2K)=\dn(z)~, \]
\[ \sn(z+2iK')=\sn(z),\, ~\cn(z+2iK')=-\cn(z), \, ~\dn(z+2iK')=-\dn(z)~.\]

\subsubsection{ MI~-~I: Case $Q=0$, $P=1$}

MI~-~I identities result if there are an odd total number of $\dn$
and $\cn$, and an even total number of $\sn$ and $\cn$ functions
in $f(z)$, i.e.
if one considers terms of the form $\dn^a~\sn^b~\cn^c$, then $a+c$ is odd
and $b+c$ is even.
The primitive function of this type is $\dn(z)$ and we consider an
``archetypal sum'' $\sigma_1$ from which identities in this class
can be derived: \beq \sigma_1(x_0)=\sum_{j=1}^{p} \mbox{dn}(x_j)~,
\eeq where $p$ is any (odd or even) integer. We note that in the
case of MI~-~I \beq a_0=\int_0^{T=2K} f(x) \, dx = i \pi
\gamma_1~, \eeq as can be seen on integrating $f(z)$ around
$ABCD$, and making use of the antisymmetry about $2iK'$, since we are considering the case $P=1$.

Since $\dn(z)$
has a single simple pole at $iK'$ interior to $ABCD$ with $\gamma_1=-i$,
using the Poisson summation formula yields
\beq \sigma_1(x_0)= \frac{p\pi}{2K}\left[ 1 \,+\, 2 \sum_{k\ne0,
p|k} \frac{ q^{-k} }{1+q^{-2k}} e\left(\f{k x_0}{2K}\right)
\right]. \eeq The above expression for $\sigma_1(x_0)$ now allows
us to re-write $S_p(x_0)$ as given in Eq. (\ref{mother1}),
yielding our first master identity:
\beq \label{master01}
S_p(x_0)= i\sum_{l=1}^{L} \frac{(-1)^{l-1}\gamma_l}{(l-1)!}
\frac{d^{l-1}}{dx_0^{l-1}} \sigma_1(x_0)~. \eeq Thus all the sums
in this class can be written as sums over the higher order
derivatives of the function $\dn(z)$. The highest derivative order
is one less than the maximum of the orders of the function $f(z)$
at all the interior poles. We see that the sums involving the
Jacobi functions are intimately related to their singularity
structure in the complex plane.

As an illustration, consider the sum
\beq
\label{ddd}
S_p(x_0)=\sum_{j=1}^p \dn(x_j)\dn(x_{j+1})\dn(x_{j+2})~.
\eeq
The relevant function is $f(z)= \dn(z)\dn(z+2K/p)\dn(z+4K/p)$, with
poles at $iK'$, $iK'-2K/p$, and $iK'-4K/p$ within $ABCD$.
The principal part of the function $\dn(z)[\dn(z+2K/p) \dn(z+4K/p)+\dn(z-2K/p)\dn(z+2K/p)+\dn(z-4K/p)
\dn(z-2K/p)]$ around $z=iK'$ determines the $\gamma_l$. The
singularity of $\dn(z)$ is simple, therefore $L=1$.
Using the identity $\dn(z+iK')=-i\,\cs(z)$ we get that
$\gamma_1=-i[\cs^2(2K/p)-2\cs(2K/p) \cs(4K/p)]~. $
Substituting this result in Eq. (\ref{master01}) gives the identity
\beq\label{ddd1}
\sum_{j=1}^p \dn(x_j)\dn(x_{j+1})\dn(x_{j+2}) \,=\,
\left[\cs^2(2K/p)-2\cs(2K/p)\cs(4K/p)\right] \sum_{j=1}^{p} \dn(x_j)~.
\eeq

As another example consider
\beq
S_p(x_0)=\sum_{j=1}^{p} \sn(x_j)\,\cn(x_j)\,\dn(x_j)[\dn(x_{j+1})+\dn(x_{j-1})]~.
\eeq
The relevant function now is $f(z)=\sn(z)\, \cn(z)\, \dn(z) [\dn(z+2K/p)+\dn(z-2K/p)]$.
There are three poles, one at $iK'$ and the others at $iK' \pm 2K/p$. To get the quantities
$\gamma_l$, it is convenient to consider the principal part of
$f(z)+f(z+2K/p)+f(z-2K/p)$ around $z=iK'$. At $iK'$, while the product of the three functions
$\sn$, $\cn$ and $\dn$ gives an order three singularity, it is reduced by one due to the
vanishing of a constant term in the expansion of $\dn(z+2K/p)+\dn(z-2K/p)$ around the
same point.
Thus we get that the maximum order of $f(z)$ is $L=2$ while
$\gamma_1=0$ and
$\gamma_2 \,=\, (-2i/m) \ds(2K/p)\, \ns(2K/p)$~.
Substitution in Eq.~(\ref{master01}) leads to the identity: \beq
\sum_{j=1}^{p}
\sn(x_j)\,\cn(x_j)\,\dn(x_j)[\dn(x_{j+1})+\dn(x_{j-1})]\,=\,
2\ds(2K/p)\, \ns(2K/p) \sum_{j=1}^{p} \cn(x_j) \,\sn(x_j)~. \eeq
Several other identities of this type are given in Appendix A.

\subsubsection{ MI~-~II: Case $Q=0$, $P=0$}

This case results when there are an even total number of $\dn$ and
$\cn$ and an even total number of $\sn$ and $\cn$ i.e. if one
considers terms of the form $\dn^a~\sn^b~\cn^c$, then both $a+c$
and $b+c$ must be even. In this case the relevant primitive
function can be taken as $\dn^2(z)$ and we define and evaluate the
following archetypal sum:
\begin{eqnarray}\label{23}
\sigma_2(x_0)=\sum_{j=1}^{p} \mbox{dn}^2(x_j)= \frac{pE}{K} +
\frac{p \pi^2}{K^2}\sum_{k\ne0,\, p|k}\frac{ k q^{k} }{1-q^{2k}}
 e\left(\f{k x_0}{2K}\right)~.
\end{eqnarray}
Here $E$ is the complete elliptic integral of the second kind
\cite{abr}. We note that in this case $\gamma_1$ is zero, as the
integral of $f(z)$ around the rectangle $ABCD$ vanishes.

Substituting Eq. (\ref{23}) in Eq. (\ref{mother1}), yields the
second master identity: \beq \label{master00} S_p(x_0)=
\frac{p}{2K}\left[ \int_0^{2K} f(x)dx + 2 \gamma_2 E \right] \, +
\sum_{l=2}^{L} \frac{(-1)^{l-1}\gamma_l}{(l-1)!}
\frac{d^{l-2}}{dx_0^{l-2}} \sigma_2(x_0)~. \eeq Thus all MI~-~II
identities have derivatives of $\dn^2(z)$ upto order $L-2$. This
is also the only master identity that has a non-vanishing
``constant'' (independent of $x_0$) term on the right hand side.
The simplest member of this class has already been discussed in
Eq.~(\ref{ddA}). We note that the relevant function for this
identity is $f(z)=\dn(z)\dn(z+2K/p)$. There are two poles, one at
$iK'$ and other at $iK'-2K/p$. Thus we can construct
$\dn(z+iK')[\dn(z+iK'+2K/p)+\dn(z+iK'-2K/p)] =
-\cs(z)[\cs(z+2K/p)+\cs(z-2K/p)]$ and its principal part around
$z=0$ will give us the $\gamma_l$. The principal part of $\cs(z)$
around $z=0$ is $1/z$. Therefore the only relevant number is
$\gamma_1=-[\cs(2K/p)+\cs(-2K/p)]=0$.  Anyway we have already
observed above that for this class $\gamma_1=0$ from the fact that
$ABCD$ is a period parallelogram for $f(z)$.  Thus the sum of the
principal parts cancel and so do all the $\gamma_l$. Hence using
Eq. (\ref{master00}), we obtain the identity~(\ref{ddB}).  In fact
we can easily generalize using the same argument to a cyclic sum
of any even number of $\dn$ or $\sn$ or $\cn$. For instance, \beq
\sum_{j=1}^{p} \dn(x_j) \dn(x_{j+r}) \dn(x_{j+s})
\dn(x_{j+t})\,=\, \f{p}{2K} \int_{0}^{2K} f(x) dx~, \eeq where
$f(x)=\dn(x)\dn(x+r2K/p)\dn(x+s2K/p)\dn(x+t2K/p)$.

As another example, we establish the identity
\beq\label{iiii}
\sum_{j=1}^{p} \dn^2(x_j)\dn^2(x_{j+1})= A \, \sum_{j=1}^{p} \dn^2(x_j)\, +\, B~.
\eeq
Writing $f(z)=\dn^2(z) \dn^2(z+2K/p)$, there are two poles of order two at
$iK'$ and $iK'-2K/p$ within $ABCD$. We find that $L=2$ and
$\gamma_2=2 \cs^2(2K/p).$
Thus applying the master identity leads to \beq
A=-2\cs^2(2K/p),\;\;
B=\f{p}{2K}\left(\int_{0}^{2K}\dn^2(t)\dn^2(t+2K/p)\, dt +4E\,
\cs^2(2K/p)\right)~. \eeq

For an example of this class with $L=3$, we prove the identity
\beq
\sum_{j=1}^{p} \cn(x_j)\,\sn(x_j)[\dn^3(x_{j+1})+\dn^3(x_{j-1})]\,=\, A\sum_{j=1}^{p} \cn(x_j)\,\sn(x_j)\,\dn(x_j)~.
\eeq
We can derive this using the master identity with $f(z)=\cn(z)\,\sn(z)[\dn^3(z+2K/p)+\dn^3(z-2K/p)]$,
 and we find that $L=3$ with $\gamma_2=0$ and
$\gamma_3\,=\, (2/m)\,\ds(2K/p)\,\ns(2K/p)$~.
Thus the first derivative of $\dn^2(z)$ will appear in the RHS, which indeed
leads to the above identity with the constant $A=-2\,\ns(2K/p) \,\ds(2K/p).$
Note that
\beq
\int_{0}^{2K}f(t)\, dt = \int_{0}^{K}[f(t)+f(-t)]\,dt =0~,
\eeq
since $f(t)$ is an odd function of $t$.

\subsection{Cases corresponding to $Q=1$}

When $Q=1$, the function $f(z)$ has a real period $4K$. We
consider the rectangle $ABCD\equiv (-\epsilon, 4K-\epsilon,
4K-\epsilon+2iK', -\epsilon+2iK')$, where $\epsilon$ is a small
positive number, and integrate around this rectangle. Poles occur
at $iK' + w4K/p$ and $iK'+2K +w4K/p$ inside the rectangle $ABCD$.
If the principal part around $iK'+w4K/p$ is given by the set of
coefficients $\{\gamma_l\}$, the set around $iK'+2K+w4K/p$ is
$\{-\gamma_l\}$, since $f(z+2K)=-f(z)$. Also note that \beq
a_0=\int_0^{4K} f(x) dx =0~, \eeq due to antisymmetry about $2K$.
Applying the Poisson summation formula and following the same
procedures as for the previous cases, we get the equivalent of Eq.
(\ref{mother1}): \beq \label{mother2} S_p(x_0)=\frac{2\pi i\, p
}{4K} \sum_{l=1}^{L} \frac{\gamma_l}{(l-1)!}\left(\frac{-2 \pi i
}{4K}\right)^{l-1}\sum_{k\ne0, p|k} \frac{k^{l-1} [1-(-1)^k]
q^{-k/2} }{1+(-1)^{P+1}q^{-k}}
 \, e\left(\f{k x_0}{4K}\right)~.
\eeq
We note that $S_{2p}(x_0)=0$, {\it i.e.} the sums in these cases
vanish for even values of $p$. This is however a trivial identity since
$f(x_j)=-f(x_{j+p/2})$ for $j=1,\ldots,p/2$. Thus, for $Q=1$, it is sufficient to only consider identities where $p$ is odd.

\subsubsection{ MI~-~III: Case $Q=1$, $P=0$}

This case applies when $f(z)$ has an even total number of $\dn$
and $\cn$ and there are an odd total number of $\sn$ and $\cn$
i.e. if one considers terms of the form $\dn^a~\sn^b~\cn^c$, then
$a+c$ is even and $b+c$ is odd. The relevant primitive function
here is $\sn(z)$ and the archetypal sum is \beq \label{32}
\sigma_3(x_0)=\sum_{i=j}^{p} \sn(x_j)= \frac{ 2 \pi i\, p}{4 K
\sqrt{m}}\sum_{k\ne0, p|k}\frac{ [1-(-1)^k] q^{-k/2} }{1-q^{-k}}
 \, e\left(\f{k x_0}{4K}\right)~.
\eeq
Therefore, using Eq. (\ref{32}) in Eq. (\ref{mother2}), we get the third master identity:
\beq
\label{master10}
S_p(x_0)= \sqrt{m}\sum_{l=1}^{L} \frac{(-1)^{l-1}\gamma_l}{(l-1)!} \frac{d^{l-1}}{dx_0^{l-1}} \sigma_3(x_0)~.
\eeq

As an illustration take $f(z)=\sn^2(z)[\sn(z+4K/p)+\sn(z-4K/p)]$. This gives
$L=1$ with
$\gamma_1=(2/m^{3/2})[\ns^2(4K/p) -\ds(4K/p)\,\cs(4K/p)]~.$
The resulting identity is
\beq
\sum_{j=1}^{p} \sn^2(x_j)[\sn(x_{j+1})+\sn(x_{j-1})]
=(2/m) \left [\ns^2(4K/p) -\ds(4K/p)\,\cs(4K/p)\right] \sum_{j=1}^{p} \sn(x_j)~.
\eeq

An example with $L=2$ is provided by $f(z)=\sn(z) \,
\dn(z)[\sn(z+4K/p) \, \cn(z+4K/p)+\sn(z-4K/p) \, \cn(z-4K/p)]$.
This results in $\gamma_1=0$ while
$\gamma_2=-(2/m^{3/2})\ns(4K/p)[\cs(4K/p)+\ds(4K/p)]$~. Therefore
the first derivative of $\sn$ will appear on the right hand side
of the identity which we write as: \beqa
&&\sum_{j=1}^{p}\sn(x_j)\,\dn(x_j) [\sn(x_{j+1}) \,
\cn(x_{j+1})+\sn(x_{j-1}) \, \cn(x_{j-1})] \nonumber \\
&&=(2/m)\ns(4K/p)[\cs(4K/p)+\ds(4K/p)]\sum_{j=1}^{p}
\cn(x_j)\,\dn(x_j)~. \eeqa

\subsubsection{ MI~-~IV: Case $Q=1$, $P=1$}

This case applies when there are an odd total number of $\dn$ and
$\cn$ and there are an odd total number of $\sn$ and $\cn$ in
$f(z)$, i.e. if one considers terms of the form
$\dn^a~\sn^b~\cn^c$, then both $a+c$ and $b+c$ must be odd. The
relevant primitive function here is $\cn(z)$ and the archetypal
sum is \beq \label{36} \sigma_4(x_0)=\sum_{i=1}^{p} \cn(x_i)=
\frac{2 \pi\,p }{4 K \sqrt{m}}\sum_{k\ne0,\, p|k}\frac{ [1-(-1)^k]
q^{-k/2} }{1+q^{-k}}\,
 e\left(\f{k x_0}{4K}\right)~.
\eeq
Therefore, using Eq. (\ref{36}) in Eq. (\ref{mother2}), we get the fourth and final master identity:
\beq
\label{master11}
S_p(x_0)= i \sqrt{m}\sum_{l=1}^{L} \frac{(-1)^{l-1}\gamma_l}{(l-1)!}
\frac{d^{l-1}}{dx_0^{l-1}} \sigma_4(x_0)~.
\eeq

As an illustration, consider  $f(z)=\cn^2(z)~[\cn(z+4K/p)+\cn(z-4K/p)]$. This gives
$L=1$ with
$\gamma_1=(2i/m^{3/2})[\ds^2(4K/p) -\ns(4K/p)\,\cs(4K/p)]~.$
The resulting identity is
\beq
\sum_{j=1}^{p} \cn^2(x_j)[\cn(x_{j+1})+\cn(x_{j-1})]=
(2/m)\left[\ns(4K/p)\,\cs(4K/p) -\ds^2(4K/p)\right] \sum_{j=1}^{p} \cn(x_j)~.
\eeq

For $f(z)=\cn^2 (z)\, \dn(z) [\sn(z+4K/p)+\sn(z-4K/p)]$, we get
$L=2$ and $\gamma_1=0$ while
$\gamma_2=(-2i/m^{3/2})\,\cs(4K/p)\,\ds(4K/p)~.$ The resultant
identity therefore involves the first derivative of $\cn$: \beq
\sum_{j=1}^{p}\cn^2 (x_j)\,\dn(x_j)[\sn(x_{j+1})+\sn(x_{j-1})]
=(2/m) \,\cs(4K/p)\ds(4K/p)\sum_{j=1}^{p} \sn(x_j)\dn(x_j)~. \eeq

This completes our enumeration of master identities for ordinary sums.
Many additional examples are given in Appendix A.

\section{Master Identities With Alternating Signs}

Alternating sums provide an immediate and important extension of
the master identities discussed above. We consider sums of the
form: \beq \label{altsum} S_p^A(x_0)=\sum_{j=1}^{p}(-1)^{j-1}
f(x_j)~. \eeq where again $f(x)$ has the properties discussed for
the ordinary sums in Eq.~(\ref{sum}). We are however forced to restrict $p$
to be {\em even} in this section and as a result
we only have MI-I and MI-II
master identities with alternating signs.
One of the consequence of
having an alternating sum, as we will see below, is the appearance
of the simply periodic Jacobi zeta function \cite{law} as an
important player.

To clarify the differences that arise between
ordinary and alternating sums we first work out an example:
\beqa\label{altex1}
S_p^A(x_0)&=&\sum_{j=1}^{p} (-1)^{j-1}
\dn^2(x_j)[\dn(x_{j+r})+\dn(x_{j-r})]\nonumber \\
&=&2\left[\ds(r2K/p)\,\ns(r2K/p)-(-1)^{r}\cs^2(r2K/p)\right]
\sum_{j=1}^{p} (-1)^{j-1} \dn(x_j)~.
\eeqa
Here the spacing is $r
2K/p$. Since $r$ and $p$ are coprimes and $p$ is restricted to be an
even integer, hence for alternating sums $r$ can only take odd integral
values.

To prove the above identity, in the case $r=1$, consider the sums $S^+_p$
and $S^-_p$, corresponding to the positive and negative signed
terms in $S^A_p$. We have to take
$f(z)=\dn^2(z)[\dn(z+T/p)+\dn(z-T/p)]$ with $T=2K$. \beq
S_p^+(x_0)=\dn^2(x_1)[\dn(x_2)+\dn(x_p)]+\dn^2(x_3)[\dn(x_4)+\dn(x_2)]+\cdots+
\dn^2(x_{p-1})[\dn(x_p)+\dn(x_{p-2})]~, \eeq and \beq
S_p^-(x_0)=\dn^2(x_2)[\dn(x_3)+\dn(x_1)]+\dn^2(x_4)[\dn(x_5)+\dn(x_3)]+\cdots+
\dn^2(x_{p})[\dn(x_1)+\dn(x_{p-1})]~. \eeq We see that
\[ S_p^-(x_0)=S_p^+(x_0+T/p)\] and
\beq
S_p^+(x_0)=\sum_{j=1}^{\ti{p}} \dn^2[x_0+jT/\ti{p}]\left[\dn(x_0+jT/\ti{p}+T/p)
+\dn(x_0+jT/\ti{p}-T/p\right)]~,
\eeq
where we have defined $\ti{p}=p/2$. The important point to note is
that while the above sum appears to be in the form of an ordinary sum
considered earlier by simply replacing $p$ with $\ti{p}$,
it is not so, as the function $f(x)$ (which usually depends on $p$) has
remained the same, or equivalently the position of the symmetric poles
is still at $iK'\pm T/p$, rather than $iK'\pm T/\ti{p}$.

Applying the Poisson summation formula we get \beq
S_p^+(x_0)=\frac{\ti{p}}{2K} \left[ {a_0} +\, \sum_{k\ne0,
\ti{p}|k} a_k \, e\left(\f{k x_0}{2K}\right) \right]~. \eeq

Note that we now need $a_k$ for $k$ that is a multiple of
$\ti{p}=p/2$ and not merely those that are multiples of $p$. For
such $k$ we get upon integrating over the same rectangle $ABCD$
that is relevant for type I ordinary identities, \beq a_k=4\pi
[\ns(2K/p)\ds(2K/p)-(-1)^{k/\ti{p}}\cs^2(2K/p)]\f{q^k}{1+q^{2k}}~.
\eeq This is because at the poles  $iK'\pm 2K/p$, $f(x)\, e\left(
k x/{2K}\right)$ has a residue of $2i \cs^2(2K/p) q^{-k}
(-1)^{k/\ti{p}}.$ Now the negative signed sum $S_p^-$ is related
to the positive signed one, by merely a shift in the argument by
an amount $2K/p$. Thus subtracting the two sums leads to a
cancellation of the zero mode term involving ${a_0}$ and also
restricts $k/\ti{p}$ to be odd integers. We then finally get \beq
S_p^A(x_0)=\f{8\pi}{2K}[\ns(2K/p)\ds(2K/p)+\cs^2(2K/p)]\sum_{k/\ti{p}=odd}\f{q^k}{1+q^{2k}}
e\left( \f{k x_0}{2K}\right)~. \eeq A similar evaluation of the
archetypal alternating sum can be done with $f(z)=\dn(z)$ which is
simpler as there is only one pole at $iK'$ within $ABCD$:
\beq\label{48}
\sigma^A_1(x_0)=\sum_{j=1}^{p}(-1)^{j-1}\dn(x_j)=\f{2\pi}{K}\sum_{k/\ti{p}
=odd}\f{q^k}{1+q^{2k}}\, e \left( \f{k x_0}{2K}\right)~. \eeq
Therefore the stated alternating identity (\ref{altex1}) follows.

We can now generalize these arguments and provide master
identities for alternating sums. Consider an elliptic function
$f(z)$ of real period $T$ satisfying Eq. (1) that has poles at
$iK'+wT/p$ where $w = 0, \pm 1,\pm 2,....$ and $|w|<p$. For both
MI-I and MI-II classes we have \beq
S^A_p(x_0)=\f{\ti{p}}{K}\sum_{k/\ti{p}=odd} a_k \, e\left(\f{k
x_0}{2K}\right)~, \eeq thus there are no constant terms, even for
type II alternating identities.

For type I and II identities we can write the $a_k$, the counterpart of
Eq.~(\ref{ordak1}) as
\begin{equation}\label{altak1}
a_k=\f{2 \pi i q^{-k} }{1+(-1)^{P+1}q^{-2k}} \sum_{l=1}^{L} \left[
\f{\ti{\gamma_l}}{(l-1)!}\left(\f{-2 \pi i }{2K}\right)^{l-1}
k^{l-1} \right]~.
\end{equation}
The difference between the two Eqs. (\ref{ordak1}) and
(\ref{altak1}) that is crucial, is that
$\ti{\gamma_l}=\sum_{w}(-1)^w\alpha_l^{(w)}$. Thus at the pole
$wT/p$, the coefficient of the order $l$ principal part gets
weighted by a factor of $(-1)^w$, as the residue calculation is restricted
to those $k$ where $k/\ti{p}$ is an odd integer. Therefore for instance
$\ti{\gamma_1}$ {\it does not in general} have the meaning of sum
of residues at all the poles. This in turn implies that it need
not vanish for type II alternating identities.

Defining the first archetypal alternating sum as in Eq. (\ref{48}), we then see
that for type I identities:
\beq
\label{masterIalt}
S^A_p(x_0)= i\sum_{l=1}^{L} \frac{(-1)^{l-1} \ti{\gamma_l}}{(l-1)!}
\frac{d^{l-1}}{dx_0^{l-1}} \sigma_1^A(x_0)~.
\eeq

Some alternating sum identities of type I are:
\beq
\label{altex3}
\sum_{j=1}^{p} (-1)^{j-1} \sn(x_j)[\cn(x_{j+1})+\cn(x_{j-1})]=0~,
\eeq
\beqa
\label{altex4}
&&\sum_{j=1}^{p} (-1)^{j-1} \sn(x_j) \,\cn(x_j)\,\dn(x_j)[\dn(x_{j+1})+\dn(x_{j-1})]
 \nonumber \\&&=2\, \ns(2K/p)\ds(2K/p)
\sum_{j=1}^{p} (-1)^{j-1} \sn(x_j) \,\cn(x_j)~.
\eeqa

Turning to the type II alternating identities, it turns out that
the primitive function in this case is Jacobi zeta function $Z(x)$
rather than $\dn^2 (x)$. To see this, consider the second
archetypal sum as \beq \sigma^A_2(x_0)=\sum_{j=1}^{p}(-1)^{j-1}
Z(x_j)~. \eeq On using the Fourier series expansion \cite{han,cop}
and the Poisson summation formula for $Z(x)$ we get: \beq
\sum_{j=1}^{p}(-1)^{j-1}Z(x_j)=\f{2 \ti{p} \pi i}{K}\sum_{k/\ti{p}
=odd}\f{q^{-k}}{1-q^{-2k}}\, e\left(\f{k x_0}{2K}\right)~. \eeq
 On
following the steps carried out above then leads to the second
master identity: \beq \label{masterIIalt} S_p^A(x_0)=
\sum_{l=1}^{L} \frac{(-1)^{l-1} \ti{\gamma_l}}{(l-1)!}
\frac{d^{l-1}}{dx_0^{l-1}} \sigma_2^A(x_0)~. \eeq

Some alternating sum identities of type II are
\beqa
\label{altex2}
&&\sum_{j=1}^{p} (-1)^{j-1} \dn(x_j)[\cn(x_{j+1})\,\sn(x_{j+1})+
\cn(x_{j-1})\,\sn(x_{j-1})] \nonumber \\
&& = -(4/m)\ds(2K/p)\,\ns(2K/p)\sum_{j=1}^{p} (-1)^{j-1} Z(x_j)~.
\eeqa
\beqa
\label{altex5}
\sum_{j=1}^{p} (-1)^{j-1} \,\cn^3(x_j)[\cn(x_{j+1})+\cn(x_{j-1})]=
(2/m^2)\cs(2K/p)\,\ns(2K/p)
\sum_{j=1}^{p} (-1)^{j-1} \dn^2(x_j)~,
\eeqa
\beqa
\label{altex6}
&&\sum_{j=1}^{p} (-1)^{j-1} \,\cn^2(x_j)\,\sn(x_j)\dn(x_j)[\cn(x_{j+1})+\cn(x_{j-1})] \nonumber \\&&
= -(4/m^2)\ds^2(2K/p)\,\cs(2K/p)\,\ns(2K/p)
\sum_{j=1}^{p} (-1)^{j-1} Z(x_j) \nonumber \\&&
+(2/m)\,\cs(2K/p)\,\ns(2K/p)
\sum_{j=1}^{p} (-1)^{j-1} \,\cn(x_j) \,\sn(x_j)\dn(x_j)~.
\eeqa

Summarizing, for functions of the form
$f(z)=h(z)[g(z+T/p)+g(z-T/p)]$, which occur in ordinary sums, we
may use the symmetrized form
$h(z)[g(z+T/p)+g(z-T/p)]+g(z)[h(z+T/p)+h(z-T/p)]$ and evaluate its
principal part at $iK'$. On the other hand, for alternating sums,
we may use the anti-symmetrized form
$h(z)[g(z+T/p)+g(z-T/p)]-g(z)[h(z+T/p)+h(z-T/p)]$ and consider its
principal part at $iK'$. Its generalization to more complex forms
of $f(z)$ is straightforward. Using the master identities derived
in this and the previous section and this methodology, we have
obtained a large number of identities, some of which are given in
Appendices A and B.

\section{Comments and Discussion}

In this section, we give some general comments and extensions in
several new directions.

\sss
{\bf \noindent(i) Identities for Auxiliary Functions:}
Until now, we have discussed identities for the three basic Jacobi elliptic functions $\sn, \cn, \dn$. However, nine auxiliary functions are also frequently found in the literature. They are:
\beqa
&&\nd\,u \equiv \frac{1}{\dn\,u}~~;~~\cd\,u \equiv \frac{\cn\,u}{\dn\,u}~~;~~\sd\,u \equiv \frac{\sn\,u}{\dn\,u}~~;\nonumber\\
&&\ns\,u \equiv \frac{1}{\sn\,u}~~;~~\cs\,u \equiv \frac{\cn\,u}{\sn\,u}~~;~~\ds\,u \equiv \frac{\dn\,u}{\sn\,u}~~;\nonumber\\
&&\nc\,u \equiv \frac{1}{\cn\,u}~~;~~\dc\,u \equiv \frac{\dn\,u}{\cn\,u}~~;~~{\rm sc}\,u \equiv \frac{\sn\,u}{\cn\,u}~~.
\eeqa

Identities for these auxiliary functions are readily obtained via the following relations \cite{abr,han}:

\beq \label{dddd}
\dn(u,m)= {\sqrt{1-m}} \,\nd\,(u-K,m)
    =-i\,\cs\,(u-iK',m)
    =i{\sqrt{1-m}}\,{\rm sc}\,(u-K-iK',m)~,
\eeq
\beq
\sn(u,m)=\,\cd\,(u-K,m)
    = \frac{1}{\sqrt{m}}\, \ns\,(u-iK',m)
    = \frac{1}{\sqrt{m}}\,\dc\,(u-K-iK',m)~,
\eeq
\beq
\cn(u,m)= -{\sqrt{1-m}} \,\sd\,(u-K,m)
= \frac{-i}{\sqrt{m}}\,\ds\,(u-iK',m)
= \frac{-i\sqrt{1-m}}{\sqrt{m}} \,\nc\,(u-K-iK',m)~.
\eeq

As an example, consider the identity $\dn(u,m)\,\dn(u+K,m)=\sqrt{1-m}$. Using Eq. (\ref{dddd}), we obtain
\beqa
\nd\,(x,m)\,\nd\,(x+K,m) &=& \frac{1}{\sqrt{1-m}} ~,\nonumber\\
\cs\,(x,m)\,\cs\,(x+K,m) &=& - {\sqrt{1-m}}~,\nonumber\\
{{\rm sc}}\,(x,m)\,{{\rm sc}}\,(x+K,m) &=& \frac{-1}{\sqrt{1-m}}~.
\eeqa

\sss {\bf \noindent(ii) Identities for Pure Imaginary Shifts:} So
far, we have focused our attention on identities involving Jacobi
elliptic functions evaluated at points separated by real gaps
$T/p$, with real $T$. As mentioned in  ref. \cite{ksjmp}, since Jacobi functions
are doubly periodic, we can convert each identity to another one
involving points separated by pure imaginary gaps $iT'/p$, with
real $T'$. The procedure consists of taking any given identity,
writing it for modulus $1-m$ [noting that $K(1-m)=K'(m)$], using
the standard results \cite{abr,han}

\beqa\label{de}
\sn(x,1-m)&=&\frac{-1}{\sqrt{1-m}}\,\dn(ix+K(m)+iK'(m),m)~,\nonumber\\
\cn(x,1-m)&=&\frac{i\sqrt{m}}{\sqrt{1-m}}\,\cn(ix+K(m)+iK'(m),m)~,\nonumber\\
\dn(x,1-m)&=&\sqrt{m}\,\sn(ix+K(m)+iK'(m),m)~,
\eeqa
and changing to a new variable $u=ix+K(m)+iK'(m)$.

For instance, again consider the simple identity $\dn(x,m)\,\dn(x+K,m)=\sqrt{1-m}$. Re-writing with modulus $1-m$, using Eq. (\ref{de}), and changing to the new variable $u=ix+K(m)+iK'(m)$ gives a simple identity involving a pure imaginary shift
\beq
\sn(u,m)\,\sn(u+iK'(m),m) = 1/\sqrt{m}~.
\eeq
A more non-trivial example consists of identity (42) in ref. \cite{ksjmp}:
\beqa
&&\sn(x,m)\,\sn(x+4K(m)/3,m)\,\sn(x+8K(m)/3,m)\nonumber\\
&&~~~~~= \frac{-1}{1-q^2}\,\left[ \sn(x,m)+\sn(x+4K(m)/3,m)+\sn(x+8K(m)/3,m)\right] ,
\eeqa
where $q \equiv \dn(2K(m)/3,m)$.

The corresponding identity with pure imaginary shifts is
\beqa
&&\dn(u,m)\,\dn(u+4iK'(m)/3,m)\,\dn(u+8iK'(m)/3,m)\nonumber\\
&&~~~~~= \frac{-(1-m)}{1-q'^2}\,\left[ \dn(u,m)+\dn(u+4iK'(m)/3,m)+\dn(u+8iK'(m)/3,m)\right] ,
\eeqa
where $q' \equiv \dn(2K'(m)/3,1-m)$.

\sss
{\bf \noindent(iii) Identities for Complex Shifts:}
Just as we have derived identities containing pure imaginary shifts, we can also derive new identities involving complex shifts. Here, the procedure consists of taking any given identity for real shifts, writing it for modulus $1/m$ (noting that $K(1/m)=\sqrt{m}[K(m)+iK'(m)]$ \cite{abr}), using the standard results
\beq
\sn(x,\frac{1}{m})=\sqrt{m}~\sn(\frac{x}{\sqrt{m}},m)~;~\cn(x,\frac{1}{m})=\dn(\frac{x}{\sqrt{m}},m)~;~\dn(x,\frac{1}{m})=\cn(\frac{x}{\sqrt{m}},m)~,
\eeq
and changing to a new variable $u=x/\sqrt{m}$.

As a simple example, let us once more take the simple identity $\dn(x,m)\,\dn(x+K,m)=\sqrt{1-m}$. It now transforms to
\beq
\cn(u,m)\,\cn(u+K(m)+iK'(m),m) = -i \frac{\sqrt{1-m}}{\sqrt{m}}~.
\eeq
As a second example, take identity (45) in ref. \cite{ksjmp}:
\beqa
&&\cn(x,m)\,\sn(x+4K/3,m)\,\sn(x+8K/3,m)+
\cn(x+4K/3,m)\,\sn(x+8K/3,m)\,\sn(x,m)\nonumber\\
&&+\cn(x+8K/3,m)\,\sn(x,m)\,\sn(x+4K/3,m)\nonumber\\
&&~~~~~~~~~= \frac{-(1+q)^2}{m}\,\left[ \cn(x,m)+\cn(x+4K/3,m)+\cn(x+8K/3,m)\right] ,
\eeqa
where $q \equiv \dn(2K(m)/3,m)$.

The corresponding identity with complex shifts is
\beqa
&&\dn(u,m)\,\sn(u+4(K+iK')/3,m)\,\sn(u+8(K+iK')/3,m)\nonumber\\
&&+~\dn(u+4(K+iK')/3,m)\,\sn(u+8(K+iK')/3,m)\,\sn(u,m)\nonumber\\
&&+~\dn(u+8(K+iK')/3,m)\,\sn(u,m)\,\sn(u+4(K+iK')/3,m)\nonumber\\
&&=-(1+r)^2\,\left[ \dn(u,m)+\dn(u+4(K+iK')/3,m)+\dn(u+8(K+iK')/3,m)\right] ,
\eeqa
where $r \equiv q(1/m)= \cn(2\{K(m)+iK'(m)\}/3,m)$.

\sss
{\bf \noindent(iv) Identities for Ratios of Jacobi Elliptic Functions:}
In applications involving linear superposition of solutions of nonlinear
differential equations \cite{ksprl,cks}, one often needs identities for
ratios of Jacobi elliptic functions like $\cn\dn/\sn$. These can be obtained
from the identities derived in this paper. For example, noting that
\beq\label{ssss}
\frac{\cn\,x~\dn\,x}{\sn\,x}=\frac{\dn\,2x\,+\,\cn\,2x}{\sn\,2x}
= i \left[\sqrt{m}\,\cn(2x+iK',m)+\dn(2x+iK',m)\right]~,
\eeq
gives rise to
\beqa\label{cccc}
&&\frac{\cn\,(x+2K/3)~\dn\,(x+2K/3)}{\sn\,(x+2K/3)}~\frac{\cn\,(x+4K/3)~\dn\,(x+4K/3)}{\sn\,(x+4K/3)}\nonumber\\
&&+\frac{\cn\,(x+4K/3)~\dn\,(x+4K/3)}{\sn\,(x+4K/3)}~\frac{\cn\,x~\dn\,x}{\sn\,x}+\frac{\cn\,x~\dn\,x}{\sn\,x}~\frac{\cn\,(x+2K/3)~\dn\,(x+2K/3)}{\sn\,(x+2K/3)}\nonumber\\
&&=-m[\cn(2u)\,\cn(2u+4K/3)+\cn(2u+4K/3)\,\cn(2u+8K/3)+\cn(2u+8K/3)\,\cn(2u)]\nonumber\\
&&-[\dn(2u)\,\dn(2u+4K/3)+\dn(2u+4K/3)\,\dn(2u+8K/3)+\dn(2u+8K/3)\,\dn(2u)]~,
\eeqa
where $u = x+iK'(m)/2$. In the above derivation, the $\cn \, \dn$ terms
cancel in view of identity (33) in ref. \cite{ksjmp}. Further, the right hand side of Eq. (\ref{cccc}) has the constant value $q(2+q)[m-(1+q)^2]/(1+q)^2~,~q \equiv \dn(2K(m)/3,m)$, due to identities (32) in ref. \cite{ksjmp}.

Other identities involving ratios follow from useful equations analogous to Eq. (\ref{ssss}):
\beqa\label{ssss1}
&&\frac{\sn\,x~\dn\,x}{\cn\,x}=\frac{1-\,\cn\,2x}{\sn\,2x}~,~
\frac{\sn\,x~\cn\,x}{\dn\,x}=\frac{1-\,\dn\,2x}{m\,\sn\,2x}~,~
\frac{\cn\,x}{\sn\,x~\dn\,x}=\frac{1+\,\cn\,2x}{\sn\,2x}~,\nonumber\\
&&\frac{\sn\,x}{\cn\,x~\dn\,x}=\frac{\dn\,2x-\,\cn\,2x}{(1-m)\,\sn\,2x}~,~
\frac{\dn\,x}{\sn\,x~\cn\,x}=\frac{1+\,\dn\,2x}{\sn\,2x}~.
\eeqa

\sss
{\bf \noindent(v) Cyclic Identities for Weierstrass Functions:}
Jacobi elliptic functions are closely related with the Weierstrass function ${\cal P}(u)$ \cite{abr,law}, the relations being:
\beq
\sn\,u=[{\cal P}(u)-e_3]^{-1/2}~,~\cn\,u=\left[\frac{{\cal P}(u)-e_1}{{\cal P}(u)-e_3}\right]^{1/2}~,~\dn\,u=\left[\frac{{\cal P}(u)-e_2}{{\cal P}(u)-e_3}\right]^{1/2}~,
\eeq
where
\beq
e_1=(2-m)/3~,~e_2=(2m-1)/3~,~e_3=-(1+m)/3~.
\eeq
${\cal P}(u)$ has implicit arguments corresponding to its two periods
$2\omega_1=2K(m)$ and $2\omega_3=2iK'(m)$ \cite{law}. Using this relationship
and identities obtained by us, we can immediately write down identities for
Weierstrass function for shifts in the units of $\omega_1 /p$, $\omega_3 /p$
and $\omega_2 /p$ where $\omega_2 =-(\omega_1 +\omega_3)$ . For example,
using identity (\ref{iiii}), one gets
\beq
\sum_{j=1}^p {\cal P}(u+2(j-1)\omega_1/p)\,{\cal P}(u+2j\omega_1/p)
=(B+pAe_1-pe_1^2)-(A-2e_1)\sum_{j=1}^p {\cal P}(u+2(j-1)\omega_1/p)
\eeq
where $A,B$ are the constants appearing in Eq. (\ref{iiii}).

\sss
{\bf \noindent(vi) Cyclic Identities for Jacobi Theta Functions:} The connection between the four Jacobi theta functions $\theta_1(z), \theta_2(z), \theta_3(z), \theta_4(z)$ and the Jacobi elliptic functions is given by \cite{law}
\beq
\sn\,u=\frac{1}{m^{1/4}}\frac{\theta_1(z)}{\theta_4(z)}~,~~
\cn\,u=\frac{(1-m)^{1/4}}{m^{1/4}}\frac{\theta_2(z)}{\theta_4(z)}~,~~
\dn\,u={(1-m)^{1/4}}\frac{\theta_3(z)}{\theta_4(z)}~,
\eeq
where $z \equiv \frac{u\pi}{2K(m)}$.
Therefore, any of our cyclic identities for real, imaginary or complex shift
can also be re-written as identities for the ratios of Jacobi elliptic
functions for shifts in units of $\pi /p$, $\pi \tau /p$ or $\pi (1+\tau)/p$
respectively where $\tau = iK'/K$.
As an illustration, we consider identity (\ref{dprod2}). In terms of theta functions, one gets
\beq
\prod_{j=1}^{p} \frac{\theta_3(z+(j-1)\pi/p)}{\theta_4(z+(j-1)\pi/p)}\,=\,
\left (\prod_{n=1}^{(p-1)/2} \frac{\theta_2^2(2nK/p)}{\theta_1^2(2nK/p)} \right) \,~
\sum_{j=1}^{p} \frac{\theta_3(z+(j-1)\pi/p)}{\theta_4(z+(j-1)\pi/p)}~.
\eeq
\sss

\sss It is a pleasure to thank Profs. Henry McKean and Ashok Raina
for insightful comments and suggestions. The work of AK and US was
supported in part by grant FGO2-84ER40173 from the U.S. Department
of Energy.

\newpage
\noindent{\bf\Large Appendix A: Examples Using the Master
Identities}\\

In this Appendix, we present a collection of identities involving
cyclic combinations of Jacobi elliptic functions. These identities
are derived by using various choices of $f(x)$ in the four master
identities [Eqs. (\ref{master01}), (\ref{master00}),
(\ref{master10}), (\ref{master11})] developed in the text. We use
the notation $a \equiv r2K/p$ and $b \equiv r4K/p$, where $1\le r
<p-1$ and $(r,p)=1$. We also use $a'=s2K/p$, $a''=t2K/p$ and
$b'=s4K/p$, where $s,r,t$ are all distinct. Also we use $s_j
\equiv \sn(x_j)$ etc., where $x_j=x_0+(j-1)T/p$. Note that $T=2K$
for the first two master identities, while it is $4K$ for the
remaining two. We note that these identities are not exhaustive
(as indeed they cannot be) but are meant to be representative low
$L$ identities.

\sss \noindent {\bf\Large Examples belonging to the class MI~-~I}

\sss\noindent $\mathbf{L=0~:}$

\beqa \sum_{j=1}^{p} s_j (c_{j+r}+c_{j-r}) \,=\,
\frac{p}{2K}\int_0^{2K}f(x)dx =0~. \eeqa

\sss\noindent $\mathbf{L=1~:}$

\beq \label{dprod1} \sum_{j=1}^{p }d_j d_{j+r}\cdots d_{j+(l-1)r}
\,=\, \left[\prod_{k=1}^{(l-1)/2}\cs^2(ka)+ 2 (-1)^{(l-1)/2}
\sum_{k=1}^{(l-1)/2}  \prod_{n=1,n \ne k}^{l} \cs([n-k]a) \right]
\sum_{j=1}^{p}d_j~, \eeq which is valid for any odd integral $l
\le p$. This is a generalization of Eq. (\ref{ddd1}) which
corresponds to $l=3$. In the special case of $l=p$ this identity
takes the simpler form \beq \label{dprod2} \prod_{j=1}^{p}
d_j\,=\, \prod_{n=1}^{(p-1)/2} \cs^2(\frac{2Kn}{p}) \,~
\sum_{j=1}^{p} d_j~. \eeq A special case of this identity for
$p=3$ has been worked out in \cite{ksjmp}.

\beq \label{ordd2d1} \sum_{j=1}^{p} d^2_j(d_{j+r}+d_{j-r}) \,=\, 2
[\ds(a) \,\ns(a) - \cs^2(a)] ~\sum_{j=1}^{p} d_j~. \eeq

\beq \sum_{j=1}^{p}
c_j(c_{j+r}d_{j+r}+c_{j-r}d_{j-r})=(-2/m)\cs(a)[\ds(a)-\ns(a)]\sum_{j=1}^{p}
d_j~. \eeq

\beq \sum_{j=1}^{p}
s_j(s_{j+r}d_{j+r}+s_{j-r}d_{j-r})=(-2/m)\cs(a)[\ds(a)-\ns(a)]\sum_{j=1}^{p}
d_j~. \eeq

\beq \sum_{j=1}^{p} d_j (d_{j+r} d_{j+s}+d_{j-r} d_{j-s})=-2\left[
\cs(a)\cs(a')+\cs(a-a')\{\cs(a)-\cs(a')\}~ \right] ~\sum_{j=1}^{p}
d_j~. \eeq

\beq \sum_{j=1}^{p} d_j (c_{j+r} c_{j+s}+ c_{j-r}c_{j-s}) =(-2/m)
\left [ \ds(a)\ds(a')+\ds(a-a')\{\cs(a)-\cs(a')\} \right]
~\sum_{j=1}^{p}d_j~. \eeq

\beq \sum_{j=1}^{p} d_j(s_{j+r} s_{j+s}+s_{j-r} s_{j-s})
=(2/m)\left[ \ns(a)\,\ns(a')+\ns(a-a')\{\cs(a)-\cs(a')\} \right]
~\sum_{j=1}^{p}d_j ~. \eeq

\beq \sum_{j=1}^{p} c_j(c_{j+r} d_{j+s}+c_{j-r}d_{j-s}) = (-2/m)
\left[ \{\cs(a')+\cs(a-a')\}\ds(a)-\ds(a-a')\ds(a')\right]
 ~\sum_{j=1}^{p}d_j~.
\eeq

\beq \sum_{j=1}^{p} s_j(s_{j+r} d_{j+s}+s_{j-r}d_{j-s}) = (2/m)
\left[\{\cs(a')+\cs(a-a')\}\ns(a)-\ns(a-a')\ns(a')\right]
~\sum_{j=1}^{p}d_j ~. \eeq

\beqa &&\sum_{j=1}^{p} s_j^2 \left [c_{j+r} c_{j+s}
d_{j+t}+c_{j-r} c_{j-s} d_{j-t} \right]=(2/m^2) \bigg
[\cs(a)\cs(a'')\ds(a')\,\ns(a) \nonumber\\
&&+\cs(a')\cs(a'')\ds(a)\,\ns(a') +
\ds(a)\ds(a')\ds(a'')\,\ns(a'')- \cs(a-a'')\ds(a-a')\,\ns^2(a)
\nonumber\\ &&-\cs(a''-a')\,\ds(a-a')\,\ns^2(a')-
\ds(a-a'')\ds(a'-a'')\,\ns^2(a') \bigg ]~\sum_{j=1}^{p}d_j ~.\eeqa

\sss\noindent $\mathbf{L=2~:}$

\beq \sum_{j=1}^{p} d^2_j[c_{j+r}s_{j+r}+c_{j-r}s_{j-r}]= -2
[\cs^2(a)+\ds(a) \,\ns(a)] ~\sum_{j=1}^{p} c_js_j~. \eeq

\beq \sum_{j=1}^{p} c_j s_j d_j [d_{j+r}+d_{j-r}]= 2
\ds(a) \,\ns(a) ~\sum_{j=1}^{p} c_js_j~. \eeq

\beq \sum_{j=1}^{p} s_j d_j(c_{j+r} d_{j+r}+c_{j-r} d_{j-r})=-2
\cs(a)[\ns(a)+\ds(a)]\sum_{j=1}^{p} c_j s_j~.\eeq

\beq \sum_{j=1}^{p} c_j (s^3_{j+r}+s^3_{j-r})=-(2/m)
\cs(a)\ns(a) \sum_{j=1}^{p} c_j s_j~.\eeq

\beq \sum_{j=1}^{p} s_j (c^3_{j+r}+c^3_{j-r})=(2/m)
\cs(a)\ds(a) \sum_{j=1}^{p} c_j s_j~.\eeq

\beq \sum_{j=1}^{p} s_j c_j(d_{j+r} d_{j+s}+d_{j-r} d_{j-s})=-2
\cs(a)\cs(a')\sum_{j=1}^{p} c_j s_j~.\eeq

\beq \sum_{j=1}^{p} s_j c_j(c_{j+r} c_{j+s}+c_{j-r}
c_{j-s})=(-2/m) \ds(a)\ds(a')\sum_{j=1}^{p} c_j s_j~.\eeq

\beq \sum_{j=1}^{p} s_j c_j(s_{j+r} s_{j+s}+s_{j-r}
s_{j-s})=(2/m)\, \ns(a)\,\ns(a')\sum_{j=1}^{p} c_j s_j~.\eeq

\beq \sum_{j=1}^{p} d_j c_j(s_{j+r} d_{j+s}+s_{j-r} d_{j-s})=-2\,
\ns(a)\,\cs(a')\sum_{j=1}^{p} c_j s_j~.\eeq

\beq \sum_{j=1}^{p} d_j s_j(c_{j+r} d_{j+s}+c_{j-r} d_{j-s})=-2\,
\ds(a)\,\cs(a')\sum_{j=1}^{p} c_j s_j~.\eeq

\beq \sum_{j=1}^{p} d^2_j (s_{j+r} c_{j+s}+s_{j-r} c_{j-s})=2\,
[\cs(a)\,\ds(a-a')-\cs(a')\ns(a-a')]\sum_{j=1}^{p} c_j s_j~.\eeq

\beqa &&\sum_{j=1}^{p} d_j c_j s_j(d_{j+r}^3 +d_{j-r}^3) \nonumber
\\ &&= -2 \left[\cs^2(a)\,\ns^2(a)+\ns^2(a)\ds^2(a)+
\ds^2(a)\cs^2(a)+3\cs^2(a)\,\ns(a)\ds(a)\right] \sum_{j=1}^{p} c_j
s_j~.\eeqa

\sss\noindent $\mathbf{L=3~:}$

\beq \sum_{j=1}^{p}
d_j^4(d_{j+r}+d_{j-r})=2\,\ns(a)\ds(a)\sum_{j=1}^{p}d_j^3+2\cs^2(a)
[\cs^2(a)-\ns(a)\ds(a)]\sum_{j=1}^{p}d_j~.\eeq

\beqa && \sum_{j=1}^{p}
d_j^3(d_{j+r}^2+d_{j-r}^2)=-2\cs^2(a)\sum_{j=1}^{p}d_j^3\nonumber
\\&&
+2\left[\cs^2(a)\,\ns^2(a)+\ns^2(a)\ds^2(a)+\ds^2(a)\cs^2(a)
-3\cs^2(a)\,\ns(a)\ds(a)\right] \sum_{j=1}^{p}d_j~.\eeqa

\beqa && \sum_{j=1}^{p}
c_j s_j d_j(c_{j+r} s_{j+r}+c_{j-r}s_{j-r})
=(2/m^2) \ds(a) \, \ns(a)\sum_{j=1}^{p}d_j^3
+(2/m) \bigg[\cs^2(a)\,\ns^2(a) \nonumber \\
&&+\ns^2(a)\ds^2(a)+\ds^2(a)\cs^2(a)
-\ns(a)\, \ds(a)(\cs^2(a)+\ns^2(a)+\ds^2(a))\bigg] \sum_{j=1}^{p}d_j~.\eeqa

\sss\noindent $\mathbf{L=4~:}$

\beq \sum_{j=1}^{p}
d_j^4(s_{j+r}c_{j+r}+s_{j-r}c_{j-r})=-2\,\ns(a)\ds(a)\sum_{j=1}^{p}s_j
c_j d_j^2+2\cs^2(a) [\cs^2(a)+3\,\ns(a)\ds(a)]\sum_{j=1}^{p}s_j
c_j ~.\eeq

\beqa && \sum_{j=1}^{p}
d_j^4(s_{j+r}c_{j+s}+s_{j-r}c_{j-s})=-2\,\ns(a)\ds(a')\sum_{j=1}^{p}s_j
c_j d_j^2\nonumber \\&&
+2\left[\cs(a)\ds(a)\cs(a')\,\ns(a')+\ns(a)\ds(a')\{\cs^2(a)+\cs^2(a')\}\right]
\sum_{j=1}^{p}s_j c_j ~.\eeqa

\sss \noindent {\bf\Large Examples belonging to the class MI~-~II}

\sss\noindent $\mathbf{L=0~:}$

\beq \sum_{j=1}^{p} d_j d_{j+r}=p \,\left( \dn(a)-\cs(a)
Z(\beta_{2rK})\right)~. \eeq

\beq \sum_{j=1}^{p} s_j s_{j+r}= \f{p}{m}\,\cs(a) Z(\beta_{2rK})~.
\eeq

\beq \sum_{j=1}^{p} c_j c_{j+r}=p \,\left( \cn(a)-\f{\ds(a)
Z(\beta_{2rK})}{m }\right)~. \eeq

It may be noted that the last two identities are valid for $p>2$
while for $p=2$ both sides vanish identically. Further, the
product of any even number ($ < p$ ) of $\dn$'s, $\sn$'s or
$\cn$'s is also a constant, the constant being the integral of the
corresponding function over the period $2K$. For example,

\beq \sum_{j=1}^{p} d_j d_{j+1} \cdots
d_{j+r-1}=\f{p}{2K}\int_{0}^{2K} \dn(x)\dn(x+2K/p) \cdots
\dn(x+(r-1)2K/p)\, dx, \;\;\mbox{for $r$ even}~.\eeq For $\dn$'s
of course even the product of all $p$ of them (i.e.
$d_1d_2...d_p$) is a constant =$(1-m)^{p/4}$ in case $p$ is even.

\beq
 \sum_{j=1}^{p} c_j d_j(s_{j+r}+s_{j-r}) = 0~.
\eeq

\beq \sum_{j=1}^{p} d_j s_j (c_{j+r}+c_{j-r}) =0~. \eeq

\beq\label{114}
 \sum_{j=1}^{p} c_j s_j(d_{j+r}+d_{j-r}) = 0~.
\eeq

\beq
 \sum_{j=1}^{p} c_j (d_{j+r} s_{j+s}+d_{j-r} s_{j-s}) = 0~.
\eeq

\beq
 \sum_{j=1}^{p} s_j (d_{j+r} c_{j+s}+d_{j-r} c_{j-s}) = 0~.
\eeq
\beq
 \sum_{j=1}^{p} d_j (c_{j+r} s_{j+s}+c_{j-r} s_{j-s}) = 0~.
\eeq

\sss\noindent $\mathbf{L=2~:}$

\beq \sum_{j=1}^{p} d_j^2 d_{j+r}^2 = A \sum_{j=1}^{p} d_j^2 +B~,
\eeq where $A=-2\cs^2(a)$, $B = \frac{p}{2K} \left(\int_0^{2K}
dn^2(t)dn^2(t+2rK/p)\,dt +4E\cs^2(a)\right)$.

\beq \sum_{j=1}^{p} c_js_j(c_{j+r}s_{j+r}+c_{j-r}s_{j-r}) =
-\gamma_2 \, \sum_{j=1}^{p} d^2_j +  \frac{p}{2K}\left(
\int_0^{2K} f(x) dx +2\gamma_2 E \right)~, \eeq where $f(x)=\cn
(x) \,\sn (x) [\cn(x+a)\,\sn(x+a)+\cn(x-a)\,\sn(x-a)]$ and
$\gamma_2=-(4/m^2)\,\ns(a) \,\ds(a)$.

\beq \sum_{j=1}^{p} c_{j}d_{j}( c_{j+r} d_{j+r}+c_{j-r}d_{j-r})
=-\gamma_2  \sum_{j=1}^{p} d^2_{j}  +
\frac{p}{2K}\left(\int_0^{2K} f(x) dx+\, 2 \gamma_2 E\right)~,
\eeq where $f(x)=\cn (x) \dn (x)[\cn(x+a) \dn(x+a)+\cn(x-a)
\dn(x-a)]$ and $\gamma_2= (4/m)\cs(a)\ds(a)$~.

\beq \sum_{j=1}^{p} s_{j}d_{j}( s_{j+r} d_{j+r}+s_{j-r}d_{j-r})
=-\gamma_2  \sum_{j=1}^{p} d^2_{j}  +
\frac{p}{2K}\left(\int_0^{2K} f(x) dx+\, 2 \gamma_2 E\right)~,
\eeq where $f(x)=\sn (x) \dn (x)[\sn(x+a) \dn(x+a)+\sn(x-a)
\dn(x-a)]$ and $\gamma_2= (-4/m)\cs(a)\,\ns(a)$~.

\beq \sum_{j=1}^{p}d_j^3(d^2_{j+r}d_{j+2r}+d^2_{j-r}d_{j-2r})
 \,=\, -\gamma_2\sum_{j=1}^{p} d^2_j
  +  \frac{p}{2K}\left(\int_0^{2K} f(x) dx+\, 2 \gamma_2 E\right)~,
\eeq where
$f(x)=\dn^3(x)[\dn^2(x+a)\dn(x+2a)+\dn^2(x-a)\dn(x-2a)]$ and
$\gamma_2 = 2 \cs(a) [ \cs^3(a)+2
\cs(2a)\ds(a)\,\ns(a)+\cs(a)\ds(2a)\,\ns(2a)]$~.

\beq \sum_{j=1}^{p}d_{j}^3( d_{j+r}+d_{j-r}) =-\gamma_2
\sum_{j=1}^{p} d^2_{j}  + \frac{p}{2K}\left(\int_0^{2K} f(x) dx+\,
2 \gamma_2 E\right)~, \eeq where $f(x)= \dn^3
(x)[\dn(x+a)+\dn(x-a)]$ and $\gamma_2= -2\,\ns(a)\ds(a)$~.

\beq \sum_{j=1}^{p}s_{j}^3( s_{j+r}+s_{j-r}) =-\gamma_2
\sum_{j=1}^{p} d^2_{j}  + \frac{p}{2K}\left(\int_0^{2K} f(x) dx+\,
2 \gamma_2 E\right)~, \eeq where $f(x)= \sn^3
(x)[\sn(x+a)+\sn(x-a)]$ and $\gamma_2= -2\cs(a)\ds(a)/m^2$~.

\beq \sum_{j=1}^{p}c_{j}^3( c_{j+r}+c_{j-r}) =-\gamma_2
\sum_{j=1}^{p} d^2_{j}  + \frac{p}{2K}\left(\int_0^{2K} f(x) dx+\,
2 \gamma_2 E\right)~, \eeq where $f(x)= \cn^3
(x)[\cn(x+a)+\cn(x-a)]$ and $\gamma_2= -2\,\ns(a)\cs(a)/m^2$~.

\beq \sum_{j=1}^{p}d_{j}^3( d_{j+r}^3+d_{j-r}^3) =-\gamma_2
\sum_{j=1}^{p} d^2_{j}  + \frac{p}{2K}\left(\int_0^{2K} f(x) dx+\,
2 \gamma_2 E\right)~, \eeq where $f(x)= \dn^3
(x)[\dn^3(x+a)+\dn^3(x-a)]$ and $\gamma_2=
12\cs^2(a)\,\ns(a)\ds(a)$~.

\beq \sum_{j=1}^{p}s_{j}^3( s_{j+r}^3+s_{j-r}^3) =-\gamma_2
\sum_{j=1}^{p} d^2_{j}  + \frac{p}{2K}\left(\int_0^{2K} f(x) dx+\,
2 \gamma_2 E\right)~, \eeq where $f(x)= \sn^3
(x)[\sn^3(x+a)+\sn^3(x-a)]$ and $\gamma_2=
-12\,\ns^2(a)\ds(a)\cs(a)/m^3$~.

\beq \sum_{j=1}^{p}c_{j}^3( c_{j+r}^3+c_{j-r}^3) =-\gamma_2
\sum_{j=1}^{p} d^2_{j}  + \frac{p}{2K}\left(\int_0^{2K} f(x) dx+\,
2 \gamma_2 E\right)~, \eeq where $f(x)= \cn^3
(x)[\cn^3(x+a)+\cn^3(x-a)]$ and $\gamma_2=
12\ds^2(a)\cs(a)\,\ns(a)/m^3$~.

\beq \sum_{j=1}^{p} c_{j} s_{j}d_{j}
(c_{j+r} s_{j+r} d_{j+r}+c_{j-r}s_{j-r}d_{j-r})
=-\gamma_2  \sum_{j=1}^{p} d^2_{j}  +
\frac{p}{2K}\left(\int_0^{2K} f(x) dx+\, 2 \gamma_2 E\right)~,
\eeq where $f(x)=\cn(x) \sn (x) \dn (x)[\cn(x+a)\sn(x+a) \dn(x+a)+\cn(x-a)
\sn(x-a) \dn(x-a)]$ and $\gamma_2= (-4/m^2)[\ns^2(a)(\cs^2(a)+\ds^2(a))
+\cs^2(a)\ds^2(a)]$~.

\sss\noindent $\mathbf{L=3~:}$

\beq \sum_{j=1}^{p} c_j s_j d_j(d_{j+r}^2+d_{j-r}^2)= -2 \cs^2(a)
\sum_{i=1}^{p} c_j s_j d_j~. \eeq

\beq \sum_{j=1}^{p} c_j s^2_j d_j(s_{j+r}+s_{j-r})= -(2/m)\,\cs(a)
\ds(a) \sum_{j=1}^{p} c_js_jd_j~. \eeq

\beq \sum_{j=1}^{p} c_j d^2_j s_j(d_{j+r}+d_{j-r})= -2 \,\ns(a)
\ds(a) \sum_{j=1}^{p} c_js_jd_j~. \eeq

\beq\label{127} \sum_{j=1}^{p} s_j c^2_j d_j(c_{j+r}+c_{j-r})=
(2/m)\,\cs(a) \,\ns(a) \sum_{j=1}^{p} c_js_jd_j~. \eeq

\beq \sum_{j=1}^{p} s_j c_j d_j^2(d_{j+r}^3+d_{j-r}^3)=
-4\,\cs^2(a) \,\ns(a) \ds(a)\sum_{j=1}^{p} c_js_jd_j~. \eeq

\beq \sum_{j=1}^{p} c_j d_j s_j^2(s_{j+r}^3+s_{j-r}^3)=
(-4/m^2)\,\ns^2(a) \ds(a) \cs(a)\sum_{j=1}^{p} c_js_jd_j~. \eeq

\beq \sum_{j=1}^{p} d_j s_j c_j^2(c_{j+r}^3+c_{j-r}^3)=
(-4/m^2)\,\ds^2(a) \cs(a) \,\ns(a)\sum_{j=1}^{p} c_js_jd_j~.\eeq

\beq \sum_{j=1}^{p} c_j s_j d_j(d_{j+r}^4+d_{j-r}^4)=
2\,\left[\cs^4(a)- \ns^2(a) \ds^2(a)-\ds^2(a)\cs^2(a)
-\cs^2(a)\,\ns^2(a)\right] \sum_{j=1}^{p} c_js_jd_j~. \eeq

\sss\noindent $\mathbf{L=4~:}$

\beqa \sum_{j=1}^{p}s_j^5(s_{j+r}+s_{j-r})&=& m^2\gamma_4
\sum_{j=1}^{p}s_j^4 +\left[m\gamma_2 -\f{2m}{3}(m+1)\gamma_4
\right]\sum_{j=1}^{p}s_j^2\nonumber \\ &&+
\f{p}{2K}\left[\int_0^{2K} f(x) dx+2\gamma_2 E\right] +p
\left(-\gamma_2+\f{m\gamma_4}{3}\right)~, \eeqa where $\gamma_2=
-(1/{3m^3})\,\cs(a)\ds(a)\left[5+5m+\cs^2(a)+\ds^2(a)+4\,\ns^2(a)\right]$,
$\gamma_4=-(2/{m^{3}})\, \cs(a)\ds(a)$ and
$f(x)=\sn^5(x)[\sn(x+a)+\sn(x-a)]$.

\beqa \sum_{j=1}^{p}d_j^4(d_{j+r}^2+d_{j-r}^2)&=& m^2\gamma_4
\sum_{j=1}^{p}s_j^4 +\left[m\gamma_2 -\f{2m}{3}(m+1)\gamma_4
\right]\sum_{j=1}^{p}s_j^2\nonumber \\
&&+\f{p}{2K}\left[\int_0^{2K} f(x) dx+2\gamma_2 E\right] +p
\left(-\gamma_2+\f{m\gamma_4}{3}\right)~, \eeqa where
$f(x)=\dn^4(x)[\dn^2(x+a)+\dn^2(x-a)]$, $\gamma_4=-2\,\cs^2(a)$
and $\gamma_2=
-(4/3)\,\cs^2(a)(m-2)-2[\cs^4(a)+\cs^2(a)\ds^2(a)+\ds^2(a)\,\ns^2(a)
+\ns^2(a)\cs^2(a)]$.

\sss \noindent {\bf\Large Examples belonging to the class
MI~-~III}

\sss\noindent $\mathbf{L=0~:}$

\beq \sum_{j=1}^{p} c_j(d_{j+r}+d_{j-r}) = 0~. \eeq

\sss\noindent $\mathbf{L=1~:}$

There are identities in $\sn$ and $\cn$ analogous to the one given
by Eq.~(\ref{ddd1}) and its generalizations in Eq.~(\ref{dprod1})
and Eq.~(\ref{dprod2}). In particular, for odd $l \le p$ we have:
\beqa &&\sum_{j=1}^{p }s_j s_{j+r}\cdots s_{j+(l-1)r} \,=\,\\
&&\f{1}{m^{(l-1)/2}}\left((-1)^{(l-1)/2}\prod_{k=1}^{(l-1)/2}\,\ns^2(kb)
+ 2 \sum_{k=1}^{(l-1)/2}  \prod_{n=1,n \ne k}^{l} \,\ns([n-k]b)
\right) ~\sum_{j=1}^{p}s_j.\nonumber \eeqa When $l=p$, the
resulting identity takes the simpler form \beq \label{sprod3}
m^{(p-1)/2}\prod_{j=1}^{p} s_j\,=\, (-1)^{(p-1)/2}\left(
\prod_{n=1}^{(p-1)/2} \,\ns^2(\frac{4Kn}{p}) \right)\,
~\sum_{j=1}^{p} s_j~. \eeq

\beq  \sum_{j=1}^{p} s^2_j(s_{j+r}+s_{j-r}) \,=(2/m)
[\ns^2(b)-\ds(b)\cs(b)] \sum_{j=1}^{p}s_j~. \eeq

\beq \sum_{j=1}^{p}
c_j(c_{j+r}s_{j+r}+c_{j-r}s_{j-r})=(2/m)\,\ns(b)[\cs(b)-\ds(b)]\sum_{j=1}^{p}
s_j~. \eeq

\beq \sum_{j=1}^{p} d_j(d_{j+r}s_{j+r}+d_{j-r}s_{j-r})=2
\,\ns(b)[-\cs(b)+\ds(b)]\sum_{j=1}^{p} s_j~. \eeq

\beq \sum_{j=1}^{p} s_j(c_{j+r} c_{j+s}+c_{j-r}c_{j-s})=
(-2/m)\left[ \ds(b)\ds(b')+\ds(b-b')\{\ns(b)-\ns(b')\} \right]
~\sum_{j=1}^{p} s_j ~. \eeq

\beq \sum_{j=1}^{p}s_j(d_{j+r} d_{j+s}+d_{j-r}d_{j-s})= -2\left [
\cs(b)\cs(b')+\cs(b-b')\{\ns(b)-\ns(b')\} \right]
~\sum_{j=1}^{p}s_j ~. \eeq

\beq \sum_{j=1}^{p}s_j(s_{j+r} s_{j+s}+s_{j-r}s_{j-s})= (2/m) \left [
\ns(b)\ns(b')+\ns(b-b')\{\ns(b)-\ns(b')\} \right]
~\sum_{j=1}^{p}s_j ~. \eeq

\beq \sum_{j=1}^{p} d_j(s_{j+r} d_{j+s}+s_{j-r}d_{j-s})= -2 \left[
\{\ns(b)-\ns(b-b')\}\cs(b') + \cs(b-b')\cs(b)\right]
 ~\sum_{j=1}^{p}s_j~.
\eeq

\beq \sum_{j=1}^{p} c_j(s_{j+r} c_{j+s}+s_{j-r}c_{j-s})= -(2/m) \left[
\{\ns(b)-\ns(b-b')\}\ds(b') + \ds(b-b')\ds(b)\right]
 ~\sum_{j=1}^{p}s_j~.
\eeq

\sss\noindent $\mathbf{L=2~:}$

\beq \sum_{j=1}^{p} c_jd_j(s^2_{j+r}+s^2_{j-r}) =
(2/m)\left[\ns^2(b) + \ds(b) \cs(b) \right]~\sum_{j=1}^{p} c_j d_j
~. \eeq

\beq \sum_{j=1}^{p} s_j d_j(s_{j+r} c_{j+r}+s_{j-r} c_{j-r})=(2/m)
\,\ns(b)\left[\cs(b)+\ds(b) \right]\sum_{j=1}^{p} c_j d_j~.\eeq

\beq \sum_{j=1}^{p} d_j c_j(s_{j+r} s_{j+s}+s_{j-r} s_{j-s})=(2/m)
\,\ns(b)\,\ns(b')\sum_{j=1}^{p} c_j d_j~.\eeq

\beq \sum_{j=1}^{p} d_j c_j(c_{j+r} c_{j+s}+c_{j-r}
c_{j-s})=(-2/m) \ds(b)\ds(b')\sum_{j=1}^{p} c_j d_j~.\eeq

\beq \sum_{j=1}^{p} d_j c_j(d_{j+r} d_{j+s}+d_{j-r} d_{j-s})=-2
\cs(b)\cs(b')\sum_{j=1}^{p} c_j d_j~.\eeq

\beq \sum_{j=1}^{p} s_j c_j(d_{j+r} s_{j+s}+d_{j-r} s_{j-s})=(2/m)
\cs(b)\,\ns(b')\sum_{j=1}^{p} c_j d_j~.\eeq

\beq \sum_{j=1}^{p} s_j d_j(c_{j+r} s_{j+s}+c_{j-r} s_{j-s})=(2/m)
\ds(b)\,\ns(b')\sum_{j=1}^{p} c_j d_j~.\eeq

\beq \sum_{j=1}^{p} c_j s_j d_j(s_{j+r}+s_{j-r})=-(2/m)
\cs(b)\,\ds(b)\sum_{j=1}^{p} c_j d_j~.\eeq

\beq \sum_{j=1}^{p} c_j (d^3_{j+r}+d^3_{j-r})=2
\cs(b)\,\ns(b)\sum_{j=1}^{p} c_j d_j~.\eeq

\beq \sum_{j=1}^{p} d_j (c^3_{j+r}+c^3_{j-r})=(2/m)
\ds(b)\,\ns(b)\sum_{j=1}^{p} c_j d_j~.\eeq

\beqa &&\sum_{j=1}^{p} d_j c_j s_j(s_{j+r}^3 +s_{j-r}^3)
= -(2/m^2) \bigg[\cs^2(b)\ns^2(b) \nonumber \\
&&+\ns^2(b)\ds^2(b)+
\ds^2(b)\cs^2(b)+3\,\ns^2(b)\cs(b)\ds(b)\bigg] \sum_{j=1}^{p} c_j
d_j~.\eeqa

\sss\noindent $\mathbf{L=3~:}$

\beqa \sum_{j=1}^{p}s_j^4(s_{j+r}+s_{j-r}) &=& -(2/m) \cs(b)
\ds(b) ~\sum_{j=1}^{p} s^3_j \nonumber \\ &+& (2/m^2)
\,\ns^2(b)[\ns^2(b)-\cs(b) \ds(b)]~\sum_{j=1}^{p} s_j~, \eeqa

\beqa &&\sum_{j=1}^{p}s_j^3(s_{j+r}^2+s_{j-r}^2) = (2/m)
\,\ns^2(b) ~\sum_{j=1}^{p} s^3_j \nonumber \\ &&+ (2/m^2)
\left[\cs^2(b)\,\ns^2(b)+\ns^2(b)\ds^2(b)
+\ds^2(b)\cs^2(b)-3\,\ns^2(b)\cs(b)\ds(b)\right]~\sum_{j=1}^{p}
s_j~. \eeqa

\beqa &&\sum_{j=1}^{p}c_j s_j d_j(c_{j+r}d_{j+r}+c_{j-r}d_{j-r}) = 2
\,\cs(b) \ds(b)~\sum_{j=1}^{p} s^3_j - (2/m)
\bigg[\cs^2(b)\,\ns^2(b) \nonumber \\
&&+\ns^2(b)\ds^2(b)
+\ds^2(b)\cs^2(b)-\,\cs(b)\ds(b)(\cs^2(b)+\ds^2(b)+\ns^2(b))\bigg]
~\sum_{j=1}^{p}
s_j~. \eeqa

\sss\noindent $\mathbf{L=4~:}$

\beqa \sum_{j=1}^{p}
&&s_j^4(d_{j+r}c_{j+r}+d_{j-r}c_{j-r})=(2/m)\cs(b)\ds(b)\sum_{j=1}^{p}c_j
d_j s_j^2 \nonumber \\ &&+(2/m^2) \,\ns^2(b)
[\ns^2(b)+3\cs(b)\ds(b)]\sum_{j=1}^{p}c_j d_j ~. \eeqa

\beqa && \sum_{j=1}^{p}
s_j^4(d_{j+r}c_{j+s}+d_{j-r}c_{j-s})=(2/m)\cs(b)\ds(b')\sum_{j=1}^{p}c_j
d_j s_j^2\nonumber \\&&
+(2/m^2)\left[\ns(b)\ds(b)\,\ns(b')\cs(b')+\cs(b)\ds(b')\{\ns^2(b)+\ns^2(b')\}\right]\sum_{j=1}^{p}c_j
d_j ~.\eeqa

\sss \noindent {\bf\Large Examples belonging to the class MI~-~IV}

\sss\noindent $\mathbf{L=0~:}$

\beq \sum_{j=1}^{p} d_j(s_{j+r}+s_{j-r}) = 0~. \eeq

\sss\noindent $\mathbf{L=1~:}$

The generalizations of Eq.~(\ref{dprod1}) and Eq.~(\ref{dprod2})
pertinent to this class are for  $l$ odd ($l \le p$): \beqa
&&\sum_{j=1}^{p }c_j c_{j+r}\cdots c_{j+(l-1)r} \,=\,\\
&&\f{1}{m^{(l-1)/2}}\left(\prod_{k=1}^{(l-1)/2}\ds^2(kb) + 2
(-1)^{(l-1)/2}\sum_{k=1}^{(l-1)/2}  \prod_{n=1,n \ne k}^{l}
\ds([n-k]b) \right) \,\sum_{j=1}^{p}c_j ~.\nonumber \eeqa When
$l=p$, the resulting identity takes the simpler form \beq
\label{cprod4} m^{(p-1)/2}\prod_{j=1}^{p} c_j\,=\,
 \prod_{n=1}^{(p-1)/2} \ds^2(\frac{4Kn}{p}) ~
 \sum_{j=1}^{p} c_j~.
\eeq

\beq \sum_{j=1}^{p}
c^2_j (c_{j+r}+c_{j-r})=(2/m) [\ns(b)\cs(b)-\ds^2(b)]\sum_{j=1}^{p}
c_j~. \eeq

\beq \sum_{j=1}^{p}
d_j(d_{j+r}c_{j+r}+d_{j-r}c_{j-r})=-2\ds(b)[\cs(b)-\ns(b)]\sum_{j=1}^{p}
c_j~. \eeq

\beq \sum_{j=1}^{p}
s_j(s_{j+r}c_{j+r}+s_{j-r}c_{j-r})=(-2/m)\ds(b)[\cs(b)-\ns(b)]\sum_{j=1}^{p}
c_j~. \eeq

\beq \sum_{j=1}^{p} d_j(c_{j+r} d_{j+s}+c_{j-r}d_{j-s})= -2\left[
\{\ds(b)-\ds(b-b')\}\cs(b')+\cs(b-b')\cs(b)\right]
~\sum_{j=1}^{p}c_j~. \eeq

\beq \sum_{j=1}^{p} s_j(c_{j+r} s_{j+s}+c_{j-r}s_{j-s})= (2/m)
\left[ \{\ds(b)-\ds(b-b')\}\,\ns(b')+\ns(b-b')\,\ns(b)\right]
~\sum_{j=1}^{p} c_j~. \eeq

\beq \sum_{j=1}^{p} c_j(s_{j+r} s_{j+s}+s_{j-r}s_{j-s})=
(2/m)\left [ \ns(b)\,\ns(b')+\ns(b-b')\{\ds(b)-\ds(b')\} \right]
~\sum_{j=1}^{p}c_j~. \eeq

\beq \sum_{j=1}^{p} c_j(d_{j+r} d_{j+s}+d_{j-r}d_{j-s})= -2 \left
[ \cs(b)\cs(b')+\cs(b-b')\{\ds(b)-\ds(b')\} \right]
~\sum_{j=1}^{p}c_j~. \eeq

\beq \sum_{j=1}^{p} c_j(c_{j+r} c_{j+s}+c_{j-r} c_{j-s})= -(2/m) \left
[ \ds(b)\ds(b')+\ds(b-b')\{\ds(b)-\ds(b')\} \right]
~\sum_{j=1}^{p}c_j~. \eeq

\sss\noindent $\mathbf{L=2~:}$

\beq \sum_{j=1}^{p} c^2_j[d_{j+r}s_{j+r}+d_{j-r}s_{j-r}]= -(2/m)
\left[\ds^2(b)+\cs(b) \,\ns(b) \right] ~\sum_{j=1}^{p} s_jd_j~.
\eeq

\beq \sum_{j=1}^{p} c_j s_j d_j [c_{j+r}+c_{j-r}]= (2/m)
\cs(b) \ns(b) ~\sum_{j=1}^{p} s_jd_j~.
\eeq

\beq \sum_{j=1}^{p} s_j c_j(d_{j+r} c_{j+r}+d_{j-r}
c_{j-r})=-(2/m) \ds(b)\left[\ns(b)+\cs(b) \right]\sum_{j=1}^{p}
s_j d_j~.\eeq

\beq \sum_{j=1}^{p} c^2_jd_j(s_{j+r}+s_{j-r})= (2/m) \cs(b)\ds(b)
~\sum_{j=1}^{p} s_j d_j ~. \eeq

\beq \sum_{j=1}^{p} d_j(s^3_{j+r}+s^3_{j-r})= -(2/m) \ds(b)\ns(b)
~\sum_{j=1}^{p} s_j d_j ~. \eeq

\beqa &&\sum_{j=1}^{p} c_j d_j s_j(c_{j+r}^3 +c_{j-r}^3)
= -(2/m^2) \bigg[\cs^2(b)\,\ns^2(b) \nonumber \\
&&+\ns^2(b)\ds^2(b)+
\ds^2(b)\cs^2(b)+3\ds^2(b)\,\ns(b)\cs(b)\bigg] \sum_{j=1}^{p} s_j
d_j~.\eeqa

\beq \sum_{j=1}^{p} s_j d_j(c_{j+r} c_{j+s}+c_{j-r}
c_{j-s})=-(2/m) \ds(b)\ds(b')\sum_{j=1}^{p} s_j d_j~.\eeq

\beq \sum_{j=1}^{p} s_j d_j(d_{j+r} d_{j+s}+d_{j-r} d_{j-s})=-2
\cs(b)\cs(b')\sum_{j=1}^{p} s_j d_j~.\eeq

\beq \sum_{j=1}^{p} s_j d_j(s_{j+r} s_{j+s}+s_{j-r} s_{j-s})=(2/m)
\,\ns(a)\,\ns(a')\sum_{j=1}^{p} s_j d_j~.\eeq

\beq \sum_{j=1}^{p} c_j d_j(s_{j+r} c_{j+s}+s_{j-r}
c_{j-s})=-(2/m) \,\ns(b)\ds(b')\sum_{j=1}^{p} s_j d_j~.\eeq

\beq \sum_{j=1}^{p} c_j s_j(d_{j+r} c_{j+s}+d_{j-r}
c_{j-s})=-(2/m) \,\cs(b)\ds(b')\sum_{j=1}^{p} s_j d_j~.\eeq

\sss\noindent $\mathbf{L=3~:}$

\beqa \sum_{j=1}^{p} s^2_jd^2_j(c_{j+r}+c_{j-r}) &=& -2 \,\ns(b)
\cs(b)\, \sum_{j=1}^{p} c^3_j  \nonumber \\
&+&(2/m)\cs(b)\,\ns^3(b)\left[m\,\sn^2(b)+\cn^2(b)-\cn(b) \right]
~\sum_{j=1}^{p} c_j ~. \eeqa

\beqa &&\sum_{j=1}^{p}c_j^3(c_{j+r}^2+c_{j-r}^2) = -(2/m) \ds^2(b)
~\sum_{j=1}^{p} c^3_j \nonumber \\ &&+ (2/m^2)
\left[\cs^2(b)\,\ns^2(b)+\,\ns^2(b)\ds^2(b)
+\ds^2(b)\cs^2(b)-3\ds^2(b)\cs(b)\,\ns(b)\right]~\sum_{j=1}^{p}
c_j~. \eeqa

\beqa &&\sum_{j=1}^{p}c_j s_j d_j (s_{j+r}d_{j+r}+s_{j-r}d_{j-r})
=2\cs(b) \ns(b)
~\sum_{j=1}^{p} c^3_j + (2/m)
\bigg[\cs^2(b)\,\ns^2(b) \nonumber \\
&&+\,\ns^2(b)\ds^2(b)
+\ds^2(b)\cs^2(b)-\cs(b)\,\ns(b)(\cs^2(b)+\ds^2(b)+\ns^2(b))\bigg]
~\sum_{j=1}^{p}
c_j~. \eeqa

\noindent{\bf\Large Appendix B: Examples Using Master Identities
With Alternating Signs}\\

Note that in this case the identities are only valid when $p$ is
even integer and since $r$ and $p$ are coprimes hence $r$ is
necessarily odd. The letters $a$, $a'$ again stand for $2rK/p$ and
$2sK/p$ respectively.

\sss \noindent {\bf\Large Examples belonging to the class MI~-~I}

\sss\noindent $\mathbf{L=1~:}$

\beqa \sum_{j=1}^{p} (-1)^{j-1} s_j (c{_{j+r}}+c{_{j-r}})=0~,
\eeqa

\beq \sum_{j=1}^{p} (-1)^{j-1} d_j d_{j+r} d_{j+2r} = -[\cs^2(a)+2
\cs(a)\cs(2a)] \sum_{j=1}^{p} (-1)^{j-1} d_j \eeq

\beq \sum_{j=1}^{p} (-1)^{j-1} d_j d_{j+r} d_{j+s} =
-\left[\cs(a)\cs(a')+ \cs(a) \cs(a'-a) - \cs(a')\cs(a'-a) \right]
\sum_{j=1}^{p} (-1)^{j-1} d_j \eeq

\beq  \sum_{j=1}^{p} (-1)^{j-1} d^2_j(d_{j+r}+d_{j-r}) \,=\, 2
[\ds(a) \,\ns(a) + \cs^2(a)] ~\sum_{j=1}^{p}(-1)^{j-1} d_j~. \eeq

\beq \sum_{j=1}^{p}(-1)^{j-1}
c_j(c_{j+r}d_{j+r}+c_{j-r}d_{j-r})=-(2/m)\cs(a)[\ds(a)+\ns(a)]
\sum_{j=1}^{p}(-1)^{j-1} d_j~. \eeq

\beq \sum_{j=1}^{p}(-1)^{j-1}
s_j(s_{j+r}d_{j+r}+s_{j-r}d_{j-r})=(2/m)\cs(a)[\ns(a)+\ds(a)]
\sum_{j=1}^{p} (-1)^{j-1}d_j~. \eeq

\beq \sum_{j=1}^{p}(-1)^{j-1}
d_j(d_{j+r}d_{j+s}+d_{j-r}d_{j-s})
=-2 [\cs(a) \cs(a') -\cs(a-a')(\cs(a)-\cs(a')]
\sum_{j=1}^{p} (-1)^{j-1}d_j~. \eeq

\beq \sum_{j=1}^{p}(-1)^{j-1}
d_j(c_{j+r}c_{j+s}+c_{j-r}c_{j-s})
=-(2/m) [\ds(a) \ds(a') -\ds(a-a')(\cs(a)-\cs(a')]
\sum_{j=1}^{p} (-1)^{j-1}d_j~. \eeq

\beq \sum_{j=1}^{p}(-1)^{j-1}
d_j(s_{j+r}s_{j+s}+s_{j-r}s_{j-s})
=-(2/m) [\ns(a) \ns(a') -\ns(a-a')(\cs(a)-\cs(a')]
\sum_{j=1}^{p} (-1)^{j-1}d_j~. \eeq

\beqa
&&\sum_{j=1}^{p}(-1)^{j-1}
s_j(s_{j+r}d_{j+s}+s_{j-r}d_{j-s}) \nonumber \\
&&=(2/m) [\ns(a) \cs(a') -\ns(a) \cs(a-a') +\ns(a') \ns(a-a')]
\sum_{j=1}^{p} (-1)^{j-1}d_j~.
\eeqa

\beqa
&&\sum_{j=1}^{p}(-1)^{j-1}
c_j(c_{j+r}d_{j+s}+c_{j-r}d_{j-s}) \nonumber \\
&&=-(2/m) [\ds(a) \cs(a') -\ds(a) \cs(a-a') +\ds(a') \ds(a-a')]
\sum_{j=1}^{p} (-1)^{j-1}d_j~.
\eeqa

\sss\noindent $\mathbf{L=2~:}$

\beq \sum_{j=1}^{p} (-1)^{j-1}
d^2_j[c_{j+r}s_{j+r}+c_{j-r}s_{j-r}]= 2[\cs^2(a)-\ds(a) \,\ns(a)]
~\sum_{j=1}^{p} (-1)^{j-1} c_js_j~. \eeq

\beq \sum_{j=1}^{p} (-1)^{j-1} s_j d_j(c_{j+r} d_{j+r}+c_{j-r}
d_{j-r})=-2 \cs(a)[-\ns(a)+\ds(a)]\sum_{j=1}^{p} (-1)^{j-1} c_j
s_j~.\eeq

\beq \sum_{j=1}^{p}(-1)^{j-1}
c_j s_j (s_{j+r}s_{j+s}+s_{j-r}s_{j-s})
=(2/m) \ns(a) \ns(a')
\sum_{j=1}^{p} (-1)^{j-1}c_j s_j~. \eeq

\beq \sum_{j=1}^{p}(-1)^{j-1}
c_j s_j (c_{j+r}c_{j+s}+c_{j-r}c_{j-s})
=-(2/m) \ds(a) \ds(a')
\sum_{j=1}^{p} (-1)^{j-1}c_j s_j~. \eeq

\beq \sum_{j=1}^{p}(-1)^{j-1}
c_j s_j (d_{j+r}d_{j+s}+d_{j-r}d_{j-s})
=-2 \cs(a) \cs(a')
\sum_{j=1}^{p} (-1)^{j-1}c_j s_j~. \eeq

\beq \sum_{j=1}^{p}(-1)^{j-1}
c_j d_j (s_{j+r}d_{j+s}+s_{j-r}d_{j-s})
=-2 \ns(a) \cs(a')
\sum_{j=1}^{p} (-1)^{j-1}c_j s_j~. \eeq

\beq \sum_{j=1}^{p}(-1)^{j-1}
s_j d_j (c_{j+r}d_{j+s}+c_{j-r}d_{j-s})
=-2 \ds(a) \cs(a')
\sum_{j=1}^{p} (-1)^{j-1}c_j s_j~. \eeq

\beq \sum_{j=1}^{p}(-1)^{j-1}
d^2_j (s_{j+r}c_{j+s}+s_{j-r}c_{j-s})
=2 [\cs(a) \ds(a-a') -\cs(a') \ns(a-a')]
\sum_{j=1}^{p} (-1)^{j-1}c_j s_j~. \eeq

\beq \sum_{j=1}^{p}(-1)^{j-1}
c_j s_j d_j (d_{j+r}+d_{j-r})
=2 \ds(a) \ns(a)
\sum_{j=1}^{p} (-1)^{j-1}c_j s_j~. \eeq

\beq \sum_{j=1}^{p}(-1)^{j-1}
s_j  (c^3_{j+r}+c^3_{j-r})
=(2/m) \cs(a) \ds(a)
\sum_{j=1}^{p} (-1)^{j-1}c_j s_j~. \eeq

\beq \sum_{j=1}^{p}(-1)^{j-1}
c_j  (s^3_{j+r}+s^3_{j-r})
=-(2/m) \cs(a) \ns(a)
\sum_{j=1}^{p} (-1)^{j-1}c_j s_j~. \eeq

\sss\noindent $\mathbf{L=3~:}$

\beqa
&&\sum_{j=1}^{p}(-1)^{j-1}
c_j s_j d_j (s_{j+r}c_{j+r}+s_{j-r}c_{j-r})
=(2/m) \ds(a) \ns(a)
\sum_{j=1}^{p} (-1)^{j-1}d^3_j \nonumber \\
&&-\bigg[\cs^2(a)\ns^2(a)+\ds^2(a)\ns^2a)+\cs^2(a)\ds^2(a)
+\ds(a)\ns(a)\big(\cs^2(a) \nonumber \\
&&+\ds^2(a)+\ns^2(a)\big)\bigg]
\sum_{j=1}^{p} (-1)^{j-1}d_j ~.
\eeqa

\sss \noindent {\bf\Large Examples belonging to the class MI~-~II}

\sss\noindent $\mathbf{L=1~:}$

While there are no ordinary identities of this class with $L=1$,
alternating identities abound. They are therefore unique and
characterized by the appearance of the Jacobian zeta function.
Also they have helped us in finding identities for the product of
$p$ $\sn$'s as well as of $p$ $\cn$'s.

\beq \sum_{j=1}^{p} (-1)^{j-1} d_jd_{j+r}=-2\cs(a) \sum_{j=1}^{p}
(-1)^{j-1}Z_j~. \eeq

\beq \sum_{j=1}^{p} (-1)^{j-1} s_js_{j+r}=(2/m)\ns(a)
\sum_{j=1}^{p} (-1)^{j-1}Z_j~. \eeq

\beq \sum_{j=1}^{p} (-1)^{j-1} c_jc_{j+r}=-(2/m)\ds(a)
\sum_{j=1}^{p} (-1)^{j-1}Z_j~. \eeq It may be noted that the above
three identities are valid for $p \ge 4$.

\beq \sum_{j=1}^{p} (-1)^{j-1} d_jd_{j+r}d_{j+2r}d_{j+3r}
=2\left[\cs(a)\cs(2a)\cs(3a)+\cs^2(a)\cs(2a)\right]~
\sum_{j=1}^{p} (-1)^{j-1}Z_j~. \eeq

This generalizes for any even number $l < p$ to:

\beq \sum_{j=1}^{p} (-1)^{j-1} d_jd_{j+r}\cdots d_{j+(l-1)r}
=2(-1)^{l/2}\left(\sum_{k=1}^{l/2} (-1)^{k-1}  \prod_{n=1,n \ne
k}^{l} \cs([n-k]a) \right) \sum_{j=1}^{p} (-1)^{j-1}Z_j~. \eeq

Similarly, for any even number $l \le p$, the $\sn$ and $\cn$
functions satisfy the identities ($p \ge 4$)

\beq m^{l/2}\sum_{j=1}^{p} (-1)^{j-1} s_js_{j+r}\cdots
s_{j+(l-1)r} =2\left(\sum_{k=1}^{l/2} (-1)^{k-1}  \prod_{n=1,n \ne
k}^{l} \ns([n-k]a) \right) \sum_{j=1}^{p} (-1)^{j-1}Z_j~. \eeq

\beq m^{l/2}\sum_{j=1}^{p} (-1)^{j-1} c_jc_{j+r}\cdots
c_{j+(l-1)r} =2(-1)^{l/2}\left(\sum_{k=1}^{l/2} (-1)^{k-1}
\prod_{n=1,n \ne k}^{l} \ds([n-k]a) \right) \sum_{j=1}^{p}
(-1)^{j-1}Z_j~. \eeq

When $l=p$ ($p \ge 4$) the above identities reduce to:

\beq
m^{p/2}\prod_{j=1}^{p}s_j=\left(\prod_{n=1}^{p/2-1}\ns^2(\frac{2Kn}{p})\right)
\sum_{j=1}^{p} (-1)^{j-1}Z_j~. \eeq

\beq
m^{p/2}\prod_{j=1}^{p}c_j=\sqrt{1-m}(-1)^{p/2}\left(\prod_{n=1}^{p/2-1}\ds^2(\frac{2Kn}{p})\right)
\sum_{j=1}^{p} (-1)^{j-1}Z_j~. \eeq

Thus these are the even product equivalents of Eqs.(\ref{sprod3})
and (\ref{cprod4}). It is worth reminding here that the product
$d_1d_2...d_p$ is on the other hand
 a constant being equal to
$(1-m)^{p/4}$.

\beq
\sum_{j=1}^{p} (-1)^{j-1} d_j [c{_{j+r}}\,s{_{j+r}}+
c{_{j-r}}\,s{_{j-r}}]= -(4/m)\ds(a)\,\ns(a)\sum_{j=1}^{p}
(-1)^{j-1} Z_j~.
\eeq
\beq
\sum_{j=1}^{p} (-1)^{j-1} s_j
[c{_{j+r}}\,d{_{j+r}}+ c{_{j-r}}\,d{_{j-r}}]=
 -(4/m)\ds(a)\,\cs(a)\sum_{j=1}^{p} (-1)^{j-1} Z_j~.
\eeq
\beq \sum_{j=1}^{p} (-1)^{j-1} c_j [s{_{j+r}}\,d{_{j+r}}+
s{_{j-r}}\,d{_{j-r}}]=
 -(4/m)\cs(a)\,\ns(a)\sum_{j=1}^{p} (-1)^{j-1} Z_j~.
\eeq

\beq
\sum_{j=1}^{p} (-1)^{j-1} d_j [c{_{j+r}}\,s{_{j+s}}+
c{_{j-r}}\,s{_{j-s}}]= (4/m)[\cs(a)\,\ds(a-a') -\cs(a')\, \ns(a-a')]
\sum_{j=1}^{p}
(-1)^{j-1} Z_j~.
\eeq

\beq
\sum_{j=1}^{p} (-1)^{j-1} s_j [d{_{j+r}}\,c{_{j+s}}+
d{_{j-r}}\,c{_{j-s}}]=-(4/m)[\ns(a)\,\ds(a-a') -\ns(a')\, \cs(a-a')]
\sum_{j=1}^{p}
(-1)^{j-1} Z_j~.
\eeq

\beq
\sum_{j=1}^{p} (-1)^{j-1} c_j [d{_{j+r}}\,s{_{j+s}}+
d{_{j-r}}\,s{_{j-s}}]=-(4/m)[\ds(a)\,\ns(a-a') -\ds(a')\, \cs(a-a')]
\sum_{j=1}^{p}
(-1)^{j-1} Z_j~.
\eeq

\sss\noindent $\mathbf{L=2~:}$

\beq \sum_{j=1}^{p} (-1)^{j-1} d_{j}^3( d_{j+r}+d_{j-r})
=2\,\ns(a)\ds(a) \sum_{j=1}^{p} (-1)^{j-1} d^2_{j}~.\eeq

\beq \sum_{j=1}^{p} (-1)^{j-1} \,c_{j}^3(c_{j+r}+c_{j-r})=
(2/m^2)\cs(a)\,\ns(a) \sum_{j=1}^{p} (-1)^{j-1} d_j^2~. \eeq

\beq \sum_{j=1}^{p} (-1)^{j-1} \,s_{j}^3(s_{j+r}+s_{j-r})=
(2/m^2)\cs(a)\,\ds(a) \sum_{j=1}^{p} (-1)^{j-1} d_j^2~. \eeq

\beqa
&&\sum_{j=1}^{p} (-1)^{j-1} \,c_{j} s_{j} d_{j}
(c_{j+r}s_{j+r}d_{j+r}+c_{j-r}s_{j-r}d_{j-r}) \nonumber \\
&&=(4/m^2)\big[\ns^2(a)(\cs^2(a)+\ds^2(a)+\cs^2(a)\ds^2(a)]
\sum_{j=1}^{p} (-1)^{j-1} d_j^2~.
\eeqa

\sss\noindent $\mathbf{L=3~:}$

\beqa &&\sum_{j=1}^{p} (-1)^{j-1}
d_j^3[c_{j+r}s_{j+r}+c_{j-r}s_{j-r}] \nonumber\\ &&=
(12/m)\cs^2(a)\ds(a)\,\ns(a) ~\sum_{j=1}^{p} (-1)^{j-1} Z_j~ -
2\,\ns(a)\ds(a)~\sum_{j=1}^{p} (-1)^{j-1} s_j c_j d_j~. \eeqa

\beqa
&&\sum_{j=1}^{p} (-1)^{j-1} \,c_{j}^2 s_{j}
d_{j}[c_{j+r}+c_{j-r}] \nonumber \\ && =
(2/m)\,\cs(a)\,\ns(a) \sum_{j=1}^{p} (-1)^{j-1} \,c_{j} s_{j}
d_{j}
-(4/m^2)\ds^2(a)\,\cs(a)\,\ns(a) \sum_{j=1}^{p} (-1)^{j-1} Z_{j}~.
\eeqa

\beqa &&\sum_{j=1}^{p} (-1)^{j-1} \,s_{j}^2 c_{j}
d_{j}[s_{j+r}+s_{j-r}] \nonumber \\ && = -(2/m)\,\cs(a)\,\ds(a)
\sum_{j=1}^{p} (-1)^{j-1} \,c_{j} s_{j} d_{j}
+(4/m^2)\ns^2(a)\,\cs(a)\,\ds(a) \sum_{j=1}^{p} (-1)^{j-1} Z_{j}.
\eeqa

\beqa &&\sum_{j=1}^{p} (-1)^{j-1} \,d_{j}^2 c_{j}
s_{j}[d_{j+r}+d_{j-r}] \nonumber \\ && = 2 \,\ds(a)\,\ns(a)
\sum_{j=1}^{p} (-1)^{j-1} \,c_{j} s_{j} d_{j}
-(4/m)\cs^2(a)\,\ds(a)\,\ns(a) \sum_{j=1}^{p} (-1)^{j-1} Z_{j}.
\eeqa

\beqa &&\sum_{j=1}^{p} (-1)^{j-1} \,d_{j} c_{j}
s_{j}[d^2_{j+r}+d^2_{j-r}] = -2 \,\cs^2(a) \sum_{j=1}^{p}
(-1)^{j-1} \,c_{j} s_{j} d_{j} \nonumber \\
&&+(4/m)[\ns^2(a)(\cs^2(a)+\ds^2(a))+\cs^2(a)\ds^2(a)]
\sum_{j=1}^{p} (-1)^{j-1} Z_{j}. \eeqa

\beqa &&\sum_{j=1}^{p} (-1)^{j-1} \,s_{j} c_{j}
[d^3_{j+r}+d^3_{j-r}] \nonumber \\ && = 2 \,\ds(a)\,\ns(a)
\sum_{j=1}^{p} (-1)^{j-1} \,c_{j} s_{j} d_{j}
-(12/m)\cs^2(a)\,\ds(a)\,\ns(a) \sum_{j=1}^{p} (-1)^{j-1} Z_{j}.
\eeqa

\beqa &&\sum_{j=1}^{p} (-1)^{j-1} \,s_{j} d_{j}
[c^3_{j+r}+c^3_{j-r}] \nonumber \\ && = (2/m) \,\cs(a)\,\ns(a)
\sum_{j=1}^{p} (-1)^{j-1} \,c_{j} s_{j} d_{j}
-(12/m^2)\ds^2(a)\,\cs(a)\,\ns(a) \sum_{j=1}^{p} (-1)^{j-1} Z_{j}.
\eeqa

\beqa &&\sum_{j=1}^{p} (-1)^{j-1} \,d_{j} c_{j}
[s^3_{j+r}+s^3_{j-r}] \nonumber \\ && = -(2/m) \,\ds(a)\,\cs(a)
\sum_{j=1}^{p} (-1)^{j-1} \,c_{j} s_{j} d_{j}
+(12/m^2)\ns^2(a)\,\ds(a)\,\cs(a) \sum_{j=1}^{p} (-1)^{j-1} Z_{j}.
\eeqa

\beqa &&\sum_{j=1}^{p} (-1)^{j-1} \,d_{j} c_{j} s^2_{j}
[s^3_{j+r}+s^3_{j-r}]  = -(8/m^2) \,\ns^2(a) \ds(a)\,\cs(a)
\sum_{j=1}^{p} (-1)^{j-1} \,c_{j} s_{j} d_{j} \nonumber \\
&&+(4/m^3)\ns^2(a)\,\ds(a)\,\cs(a) \bigg[3
\big(\ns^2(a)+\ds^2(a)+\cs^2(a) \big)+\cs^2(a)\ds^2(a) \bigg ]
\sum_{j=1}^{p} (-1)^{j-1} Z_{j}. \eeqa

\beqa &&\sum_{j=1}^{p} (-1)^{j-1} \,d_{j} s_{j} c^2_{j}
[c^3_{j+r}+c^3_{j-r}]  = -(8/m^2) \,\ds^2(a) \ns(a)\,\cs(a)
\sum_{j=1}^{p} (-1)^{j-1} \,c_{j} s_{j} d_{j} \nonumber \\
&&+(4/m^3)\ns(a)\,\cs(a) \bigg [3\ds^2(a)
\big(\ns^2(a)+\ds^2(a)+\cs^2(a) \big )+\cs^2(a)\ns^2(a) \bigg ]
\sum_{j=1}^{p} (-1)^{j-1} Z_{j}. \eeqa

\beqa &&\sum_{j=1}^{p} (-1)^{j-1} \,c_{j} s_{j} d^2_{j}
[d^3_{j+r}+d^3_{j-r}]  = -8 \,\cs^2(a) \ns(a)\,\ds(a)
\sum_{j=1}^{p} (-1)^{j-1} \,c_{j} s_{j} d_{j} \nonumber \\
&&+(4/m)\ns(a)\,\ds(a) \bigg [3\cs^2(a)
\big(\ns^2(a)+\ds^2(a)+\cs^2(a) \big )+\ds^2(a)\ns^2(a) \bigg ]
\sum_{j=1}^{p} (-1)^{j-1} Z_{j}. \eeqa

\end{document}